\documentclass[preprint,3p,twocolumn]{elsarticle}
\usepackage{graphicx}
\usepackage{mathrsfs}
\usepackage{algorithm,algorithmicx, algpseudocode}
\usepackage{amsmath}
\usepackage{caption}
\usepackage{multicol}
\usepackage[export]{adjustbox}
\usepackage{lineno,hyperref}
\usepackage{fancyvrb}

\modulolinenumbers[5]

\journal{Journal of \LaTeX\ Templates}

\begin{document}

\begin{frontmatter}

\title{A fixed storage distributed graph database hybrid with at-scale OLAP expression and I/O support of a relational DB: Kinetica-Graph$^{\vcenter{\hbox{\includegraphics[height=0.35cm]{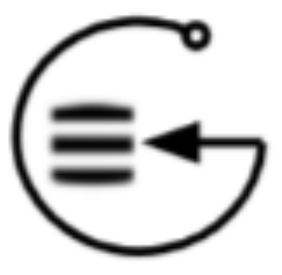}}}}$}

\author{B. Kaan Karamete\corref{cc}} 
\author{Louai Adhami\corref{}}
\author{Eli Glaser\corref{}}

\cortext[cc]{\it Corresponding author: B. Kaan Karamete,~~kkaramete@kinetica.com}
\address{Kinetica DB Inc. \break 901 North Glebe Road, Arlington, Virginia 22203}

\begin{abstract}
A distributed graph database architecture that co-exists with the distributed relational DB for I/O and at-scale OLAP expression support with hundreds of PostGIS compatible geometry functions will be discussed in this article. The uniqueness of this implementation stems mainly from its double link topology structure for its fixed storage characteristics independent from the variance in node-to-edge connections. Another note-worthy contribution of this implementation is its non-blocking client-server communication architecture among its distributed graph servers. A non-bottlenecking partitioning scheme based on duplication of nodes is also implemented ensuring minimal communications using distributed filtering on geo-spatial, random and explicit sharding choices. Finally, an efficient re-balancing algorithm followed by a distributed shortest path solver will be demonstrated with examples from both geo-spatial and social networks. 

\noindent 

\end{abstract}

\begin{keyword}
\it Graph Databases, Edge Connections, Distributed Network Solvers, Partitioned Graphs
\end{keyword}

\end{frontmatter}

\section{Introduction}

The tabular form of the data in relational databases has been the work horse of the transactional organizations for many decades. Even today, it is arguably the  most preferred method of storing and updating the data and running analytic queries over it. The table format of relational databases (relational DBs) constituted as rows and columns in this sense, much referred to as structured data, are used to compare and join with other tables by matching the records over primary and foreign keys, the unique identifiers across rows. The tabular nature of the data allowed the partitioning of the data in fixed sized fragments often depicted and known as columnar chunk format. Chunked data in columnar order is easily distributed across nodes of a cluster, or over the resources of a cloud provisioning as of late, and conveniently cast to formulate the matrix based join operations to be performed in parallel. In other words, both process and data parallelism is achieved by slicing the structured data as columnar chunks.\cite{oracle, memsql, hadoop, terradata, microsoft, kinetica}. 

During the course of the evolution in relational DB technology, various parallel formulations of the matrix based join and filter operations over the stenciled chunked data have been implemented by many relational DB vendors.\cite{kinetica,omni,snowflake}. The speed of computing queries in this manner, is often based on how effective  the data is  pulled from different storage media. Over the decades since 1970s, there has been a gradual increase of transferring data from disk to memory where the computation of analytic queries occur. The flexibility (elasticity) in adjusting the amount of the data transferring from disk, over to memory (tiering) for the duration of computation is the ultimate success criteria considering the concurrent nature of the queries. 

Individual records (row-wise) can be related to the other records via their columnar values in structured data. Hence, the comparison across these columnar content from different tables (blobs) is a squared relation-ship and therefore inevitably global. There are, however, ways to reduce the global aspect to a more local or narrower relationship among records. Data can be partitioned over a range (window-functions) so that only the portion within the range is used for generating stencils instead of the entirety of the table. Data can also be hashed in a way to skip (chunk-skipping) certain sections for generating  stencils only where it matters. However, both of these mitigation techniques require prior knowledge on the content of the data, so that certain partitioning and/or hashing (sharding) schemes can be prescribed in storing the data that will also have a huge impact on the speed of processing, say, when it is required to join with the other tables~\cite{oracle, microsoft, kinetica, windowfunctions}.

\begin{figure}
\centering
    \includegraphics[width=\linewidth, keepaspectratio]{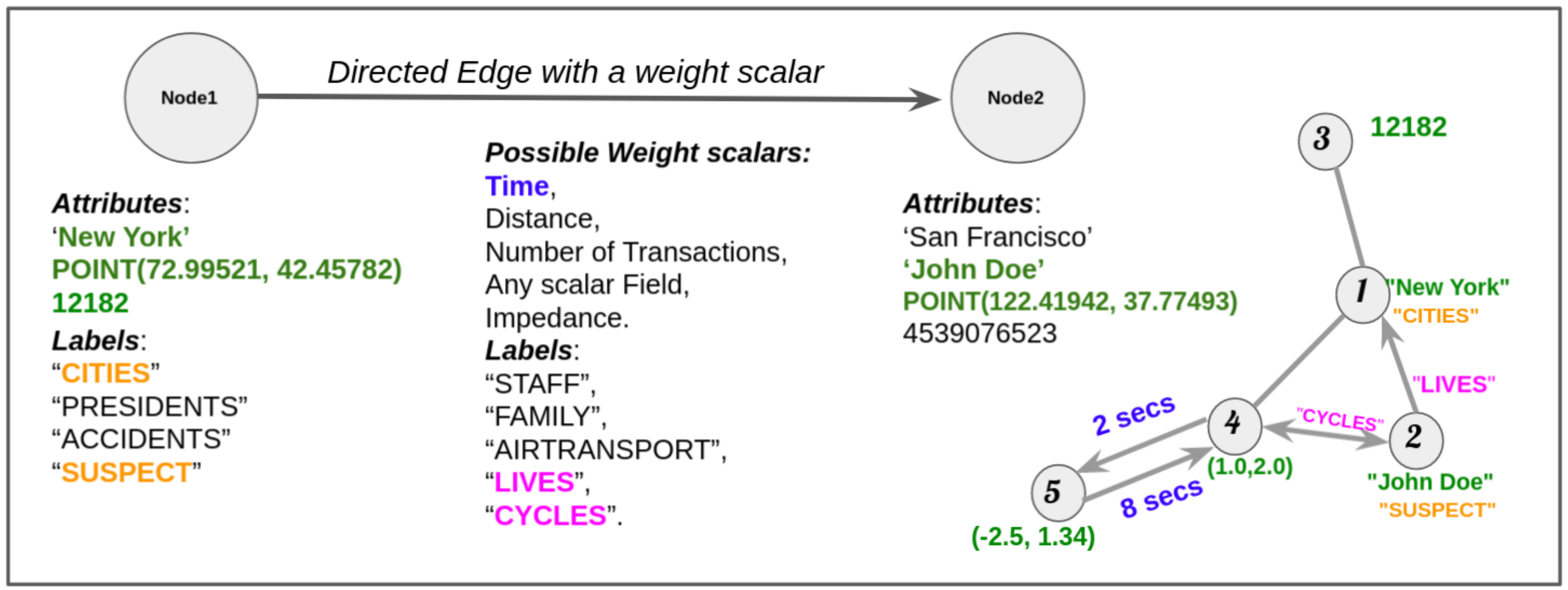}
    \caption{A typical graph layout with labels and scalar weights on nodes and edges.}
    \label{graphdb}
\end{figure}

Unstructured data, on the other hand, enables traversing the queries  in the closure of nodal relations. Nodes and pairwise connected nodes, i.e., edges can also be created from the structured data\cite{kinetica,graphdbs}, hence, establishing how the nodes are connected gives us the ability to jump across the square nature of matrix based table of records. Generally speaking, graphs are generated from these node to edge topology connections~\cite{kinetica,tiger,neo4j}. However, this has an important issue in that the other existing but unused columns should be associated with the nodes and edges as attributes. Mapping the attribute rich structured data forms to the data concise unstructured nodes and edges of a graph is usually done by using other columns as associative links as labels (See Figure~\ref{graphdb}). As in any translation, this would either result in loss of data due to the ad-hoc choice of keeping what should be on the graph or unnecessary data duplication\cite{graphdbs, tiger, neo4j, graphx}. Our graph architecture eliminates this problem by keeping the attribute rich columns at relational DB and still be able to do restrictions during graph traversals by filter expressions using the already distributed parallel OLAP engine. 

Another major issue, which is a common problem in any mesh-like structure is the dynamically changing node to edge connections, i.e., there can be varying number of edges emanating from each node, and in a dynamic table update scenario where new records can be inserted or removed, these growing and shrinking edges to an existing node can result in de-fragmentation and reallocation issues, both of which amount to excessive and often prohibitive storage requirements. Therefore, design of the graph topology in tackling this issue is one of the most essential in the efficiency of any graph database and this will be addressed in Section~\ref{Section:topology}. Unlike the other graph engines, such as Tiger and Neo4j available today~\cite{tiger,neo4j}, that have either the explicit data duplication via file I/O or hooks streaming data from third party databases like Postgres~\cite{postgres}, Kinetica-Graph co-exists with the distributed Kinetica-DB as a hybrid graph data-structure. In Section~\ref{Section:hybrid}, the details of this hybrid design will be discussed to demonstrate the effectiveness of combining the parallel OLAP computation engine using a network agnostic grammar.  

Using multiple graph servers (processors) on a cluster of machines has two benefits; it enables processing the tasks faster by creating identical copies of the graph onto each server and dividing the input accordingly among the servers (replicated) and secondly it can distribute big graphs into manageable sized sub-graphs in each server (partitioned). These two benefits, either by replicating or partitioning the graph both require non-bottlenecking communications not only among graph servers but also with the co-existing relational DB. Distributed graph servers with non-blocking communication will be discussed in Section~\ref{Section:communication} using an efficient socket communication library ZeroMQ~\cite{zmq}. 

Partitioning of graphs in Kinetica is designed to minimize the communications between its graph servers sharing the portions of a big graph as partitioned sub-graphs with only interface nodes being duplicated so that no intra-processor book keeping would be required for simplicity. There are many efficient parallel partitioning libraries in the literature that set-up and use data in their own format and communication patterns such as Zoltan and Parmetis~\cite{zoltan, parmetis}. Kinetica has chosen a more direct approach due to the fact that its data is already distributed in its hybrid relational DB. Basic partitioning filters are implemented using OLAP expressions that employ geometric, random and user provided external partitioning criteria, followed by a novel re-balancing algorithm that uses the iso-cost levels of an unbalanced distributed shortest path solve, which will be discussed in Section~\ref{Section:rebalancing}. finally, a distributed shortest path solving algorithm with examples from both geo-spatial and social networks will be explained and demonstrated in Section~\ref{Section:distributed}, respectively.

\section{Graph topology}
\label{Section:topology}

Graph storage in Kinetica is a fixed amount independent from how dense or variant the node-to-edge connections are. For a billion node graph where each node goes to every other, the conventional graph data structures require hundred million GBytes (formidable) storage whereas Kinetica Graph requires only ten GBytes. Key differentiator of Kinetica Graph DB is its efficient data representation supporting a very large number of edges/nodes that has no memory degradation under dynamic updates~\cite{dls, mapmatching}. Our optimized parallel graph solvers are built on top of this representation. Even a single node graph server can accommodate multi-billion edge graphs with dynamic upserts  streaming from DB tables easily with fixed storage characteristics that scales linearly with the graph size. One of the conventional graph edge data structures is the use of CSR format\cite{gunrock, networkx}, which is a static data structure as it keeps the start and the end node indexes of each edge in just one vector. However, either deleting an edge or adding a new one requires this vector to be updated resulting in huge reallocation blocks to be shifted, in order to insert or take out the relevant sections in the CSR vector. There is a rather quick fix for solving this issue by tombing (tagging instead of deleting) the deleted edges by simply holding an extra bit-wise information per edge, however, this approach is only good for the entity deletions, and at some point it would require a global compactification process on the entire data structure. Nevertheless, as long as there are no modifications, static CSR format has also its advantages, since node to edge iterations almost always remain within ranges of the fast $L1$ and $L2$ caches~\cite{memory}. 

\begin{figure*}
\centering
    \includegraphics[height=0.3\textheight,width=0.7\linewidth, keepaspectratio]{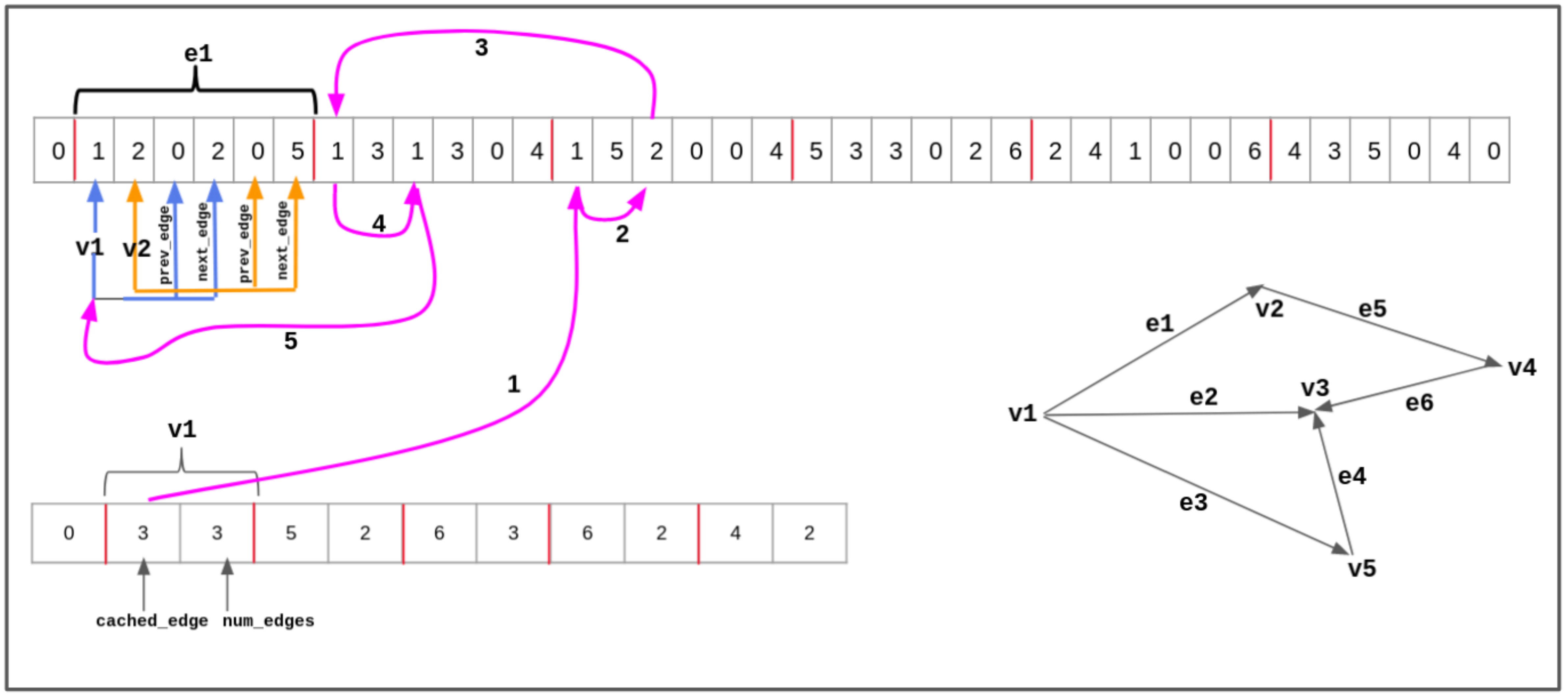}
    \caption{Double Link Graph Edge Topology: each edge record has six values; a pair of nodes, and their corresponding previous and next edge indexes - Traversal of finding upward edge links to vertex $1$: Vertex $1$ keeps a cached edge id of $3$, traversal goes to edge vector at index $3$, finds the vertex $1$, and iterates to its corresponding previous edge index of $2$, and this iteration is repeated until hitting null entry (zero index). The result is edges ${1,2,3}$ that are adjacent to vertex $1$ }
    \label{dls}
\end{figure*}

\begin{figure}
\centering
    \includegraphics[width=\linewidth, keepaspectratio]{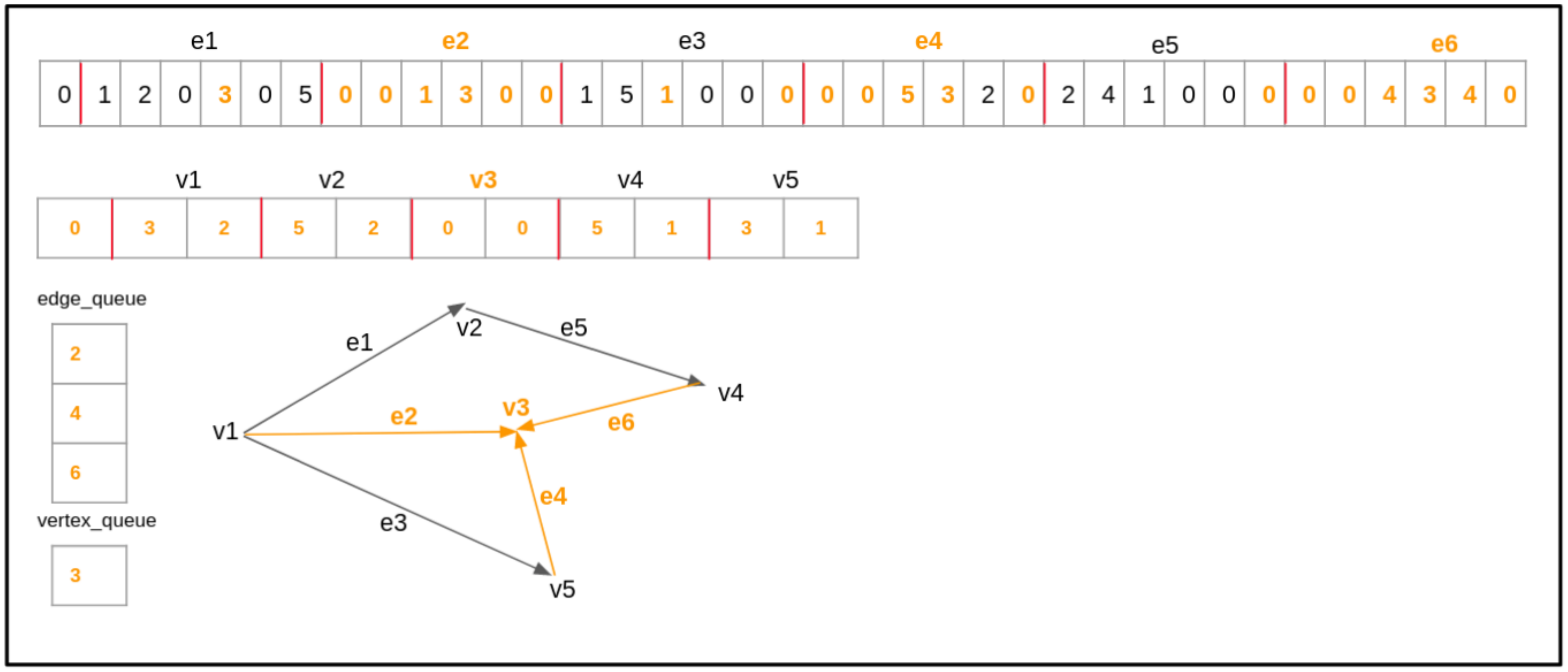}
    \caption{Recycling edge and node ids, when deletion happens; the recycled ids are used for the new node and edge insertions, hence there is no wasted space in the double link topology vectors: Vertex $3$ is deleted and edge indexes, ${2,4,6}$ are stored in a queue for reuse.}
    \label{recycleids}
\end{figure}

Our graph data structure is elegantly solving the issue of dynamic sizing of node-to-edge connections by linking edges to each other through the two nodes of each edge via the previous and the next edge links. Each node of an edge, has a previous and next edge indexes, so that we can unravel the edges of a node starting from an already cached edge index at the same node. This data structure, namely, double link structure,  DLS, devised by Karamete~\cite{dls, hexdom} on mesh based structures effectively applied on the graph topology containers at Kinetica. The novel idea is illustrated in Figure~\ref{dls}. The amount of storage is a pre-computed fixed amount of six times the number of edges; two edge links per node pair of an edge, i.e., $e_i={2 (nodes)+ 2 (previous) + 2(next)}$. The only downside of this choice is that iterating the edges of a node requires jumping along the one edge-node vector which can lead to $L1$ and $L2$ cache misses if movement range widens. One easy solution to this issue is the conversion to a CSR format temporarily if/when the graph data is static and small, though without the conversion, the delay on the speed for a typical shortest path solve is tolerably low (less than $\approx~10\%$). We have also adopted a similar tombing strategy in our DLS implementation by reusing the ids of the deleted entities to further reduce memory consumption for the newly inserted nodes/edges as depicted in Figure~\ref{recycleids}. 

\section{Hybrid graph grammar}
\label{Section:hybrid}

\begin{figure}
\centering
    \includegraphics[width=\linewidth]{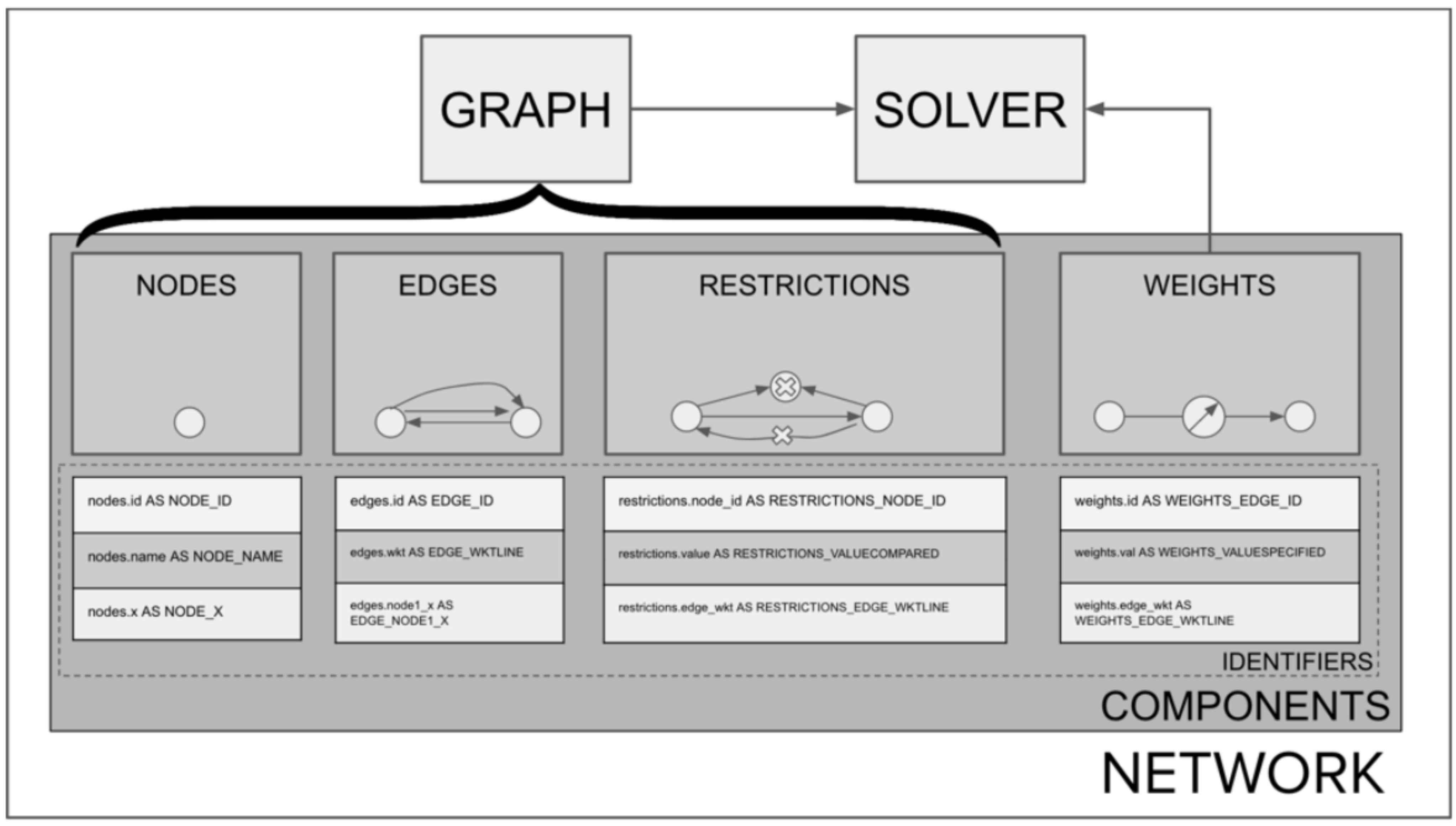}
    \caption{Kinetica-Graph network elements: components as nodes, edges, weights and restrictions, depicted via a set of ad-hoc network agnostic graph grammar identifiers, constituting the network combined with solvers.}
    \label{components}
\end{figure}

Kinetica Graph is designed and implemented from scratch driven via an extend-able and intuitive set of robust graph grammar defined as components and identifiers as annotations to DB table columns, as well as string based node and edge LABELS. The key idea behind having a network agnostic graph grammar is that nodes, edges and attributes for labels can all be transformed into graph topologies and associative maps in a unified manner regardless of the network type. The components are defined as {\it NODES, EDGES, WEIGHTS}, and {\it RESTRICTIONS} as shown in Figure~\ref{components}. A set of ad-hoc identifiers is created for each component, that can be constructed by more than one manner via a set of pre-defined tuples of these identifier combinations. For instance, a component identifier combination set of {\it (EDGE\_ID, EDGE\_NODE1\_NAME, EDGE\_NODE2\_NAME)} can be used to construct graph edges by associating each of these identifiers with a DB table column such as seen in Table~\ref{edge_combinations}. 

The geographical Lon/Lat coordinates in terms of WKT Points and LineStrings can directly be consumed by the geo-graph creation end-point utilizing an uniform bin hashing technique with a user controlled merge tolerance (graph decimation). All the heavy lifting is done by the graph server, and the only user requirement is to annotate table columns with the appropriate identifier combinations. See Tables~\ref{node_identifiers}, and~\ref{edge_identifiers}, for select number of node and edge component identifiers, respectively. 

\begin{table}[ht]
\tiny
\caption{Selected Node Component Identifiers} 
\begin{tabular}{ll rr} 
\hline\hline 
\\ [0.2ex]
&NODE\_ID& node's integer id.  \\ 
\hline \\
&NODE\_X	&node's longitude or X. \\ 
\hline \\
&NODE\_Y	&node's latitude or Y. \\ 
\hline \\
&NODE\_NAME	&node's name. \\ 
\hline \\
&NODE\_WKPOINT	&node's wktpoint \\
& &\hspace{0.125\linewidth}'POINT(-77.3808 38.7567)' \\ 
\hline \\
&NODE\_LABEL &node's label. \\ 
\hline 
\end{tabular}
\label{node_identifiers} 
\end{table}

\begin{table}[ht]
\tiny
\caption{Selected Edge Component Identifiers} 
\begin{tabular}{ll rr} 
\hline\hline 
\\ [0.2ex]
&EDGE\_ID	&edge's integer id\\ 
\hline \\
&EDGE\_NODE1\_ID	&edge's $1^{st}$ node id\\ 
\hline \\
&EDGE\_NODE2\_ID	&edge's $2^{nd}$ node id\\ 
\hline \\
&EDGE\_NODE1\_NAME	&edge's $1^{st}$ node name\\ 
\hline \\
&EDGE\_NODE2\_NAME	&edge's $2^{nd}$ node name\\ 
\hline \\
&EDGE\_DIRECTION	&edge's direction\\
\hline \\
&EDGE\_LABEL	&edge's label\\ [0.1ex]
\hline \\
&EDGE\_WEIGHT\_VALUESPECIFIED	&edge's weight\\ [0.1ex]
\hline \\
&EDGE\_WKTLINE	&edge's wktlinestring \\
\hline 
\end{tabular}
\label{edge_identifiers} 
\end{table}

\begin{table}[ht]
\tiny
\caption{Selected Edge Identifier Combinations} 
\begin{tabular}{ll} 
\hline\hline 
\\ [0.2ex]
&EDGE\_ID, EDGE\_NODE1\_ID, EDGE\_NODE2\_ID \\ 
\hline \\
&EDGE\_ID, EDGE\_NODE1\_ID, EDGE\_NODE2\_ID, \\
&\hspace{0.145\linewidth} EDGE\_DIRECTION \\
\hline \\
&EDGE\_ID, EDGE\_NODE1\_NAME, EDGE\_NODE2\_NAME \\
\hline \\
&EDGE\_ID, EDGE\_NODE1\_WKTPOINT, \\
&\hspace{0.145\linewidth} EDGE\_NODE2\_WKTPOINT \\
\hline \\
&EDGE\_ID, EDGE\_WKTLINE \\
\hline \\
&EDGE\_ID, EDGE\_WKTLINE, EDGE\_DIRECTION \\
\hline \\
&EDGE\_NODE1\_ID, EDGE\_NODE2\_ID \\
\hline \\
&EDGE\_NODE1\_NAME, EDGE\_NODE2\_NAME \\
\hline \\
&EDGE\_NODE1\_NAME, EDGE\_NODE2\_NAME, EDGE\_LABEL \\
\hline \\
&EDGE\_NODE1\_WKTPOINT, EDGE\_NODE2\_WKTPOINT \\
\hline \\
&EDGE\_WKTLINE \\
\hline \\
&EDGE\_WKTLINE, EDGE\_DIRECTION \\
\hline 
\end{tabular}
\label{edge_combinations} 
\end{table}

\begin{figure}
\centering
    \includegraphics[height=0.3\textheight, keepaspectratio]{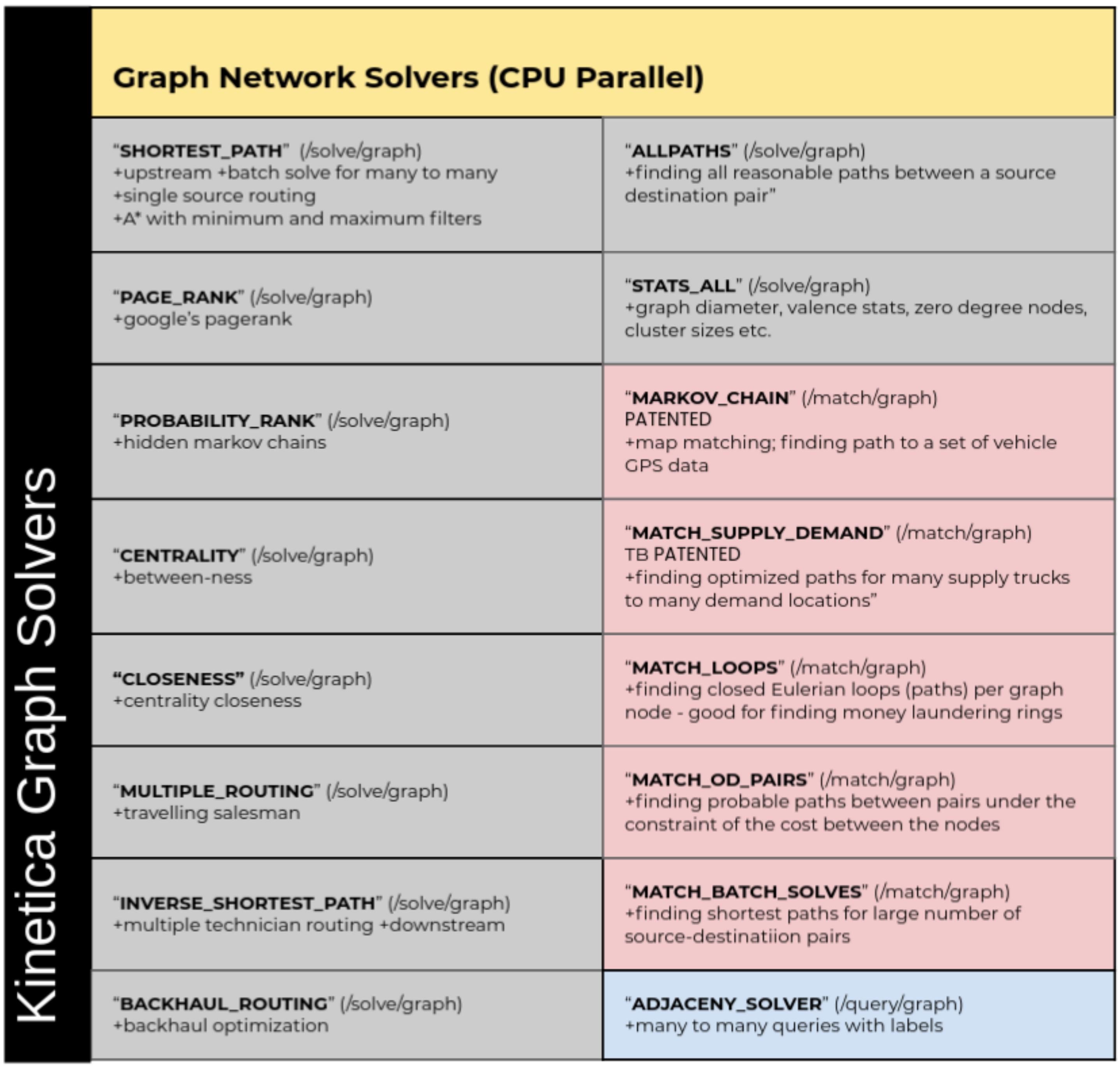}
    \caption{Kinetica-Graph Solvers: solvers exposed via solve/graph, match/graph, and query/graph apis are depicted in pale gray, pink and cyan, respectively.}
    \label{solvers}
\end{figure}

\begin{figure*}
\centering
    \includegraphics[width=0.6\textheight, keepaspectratio]{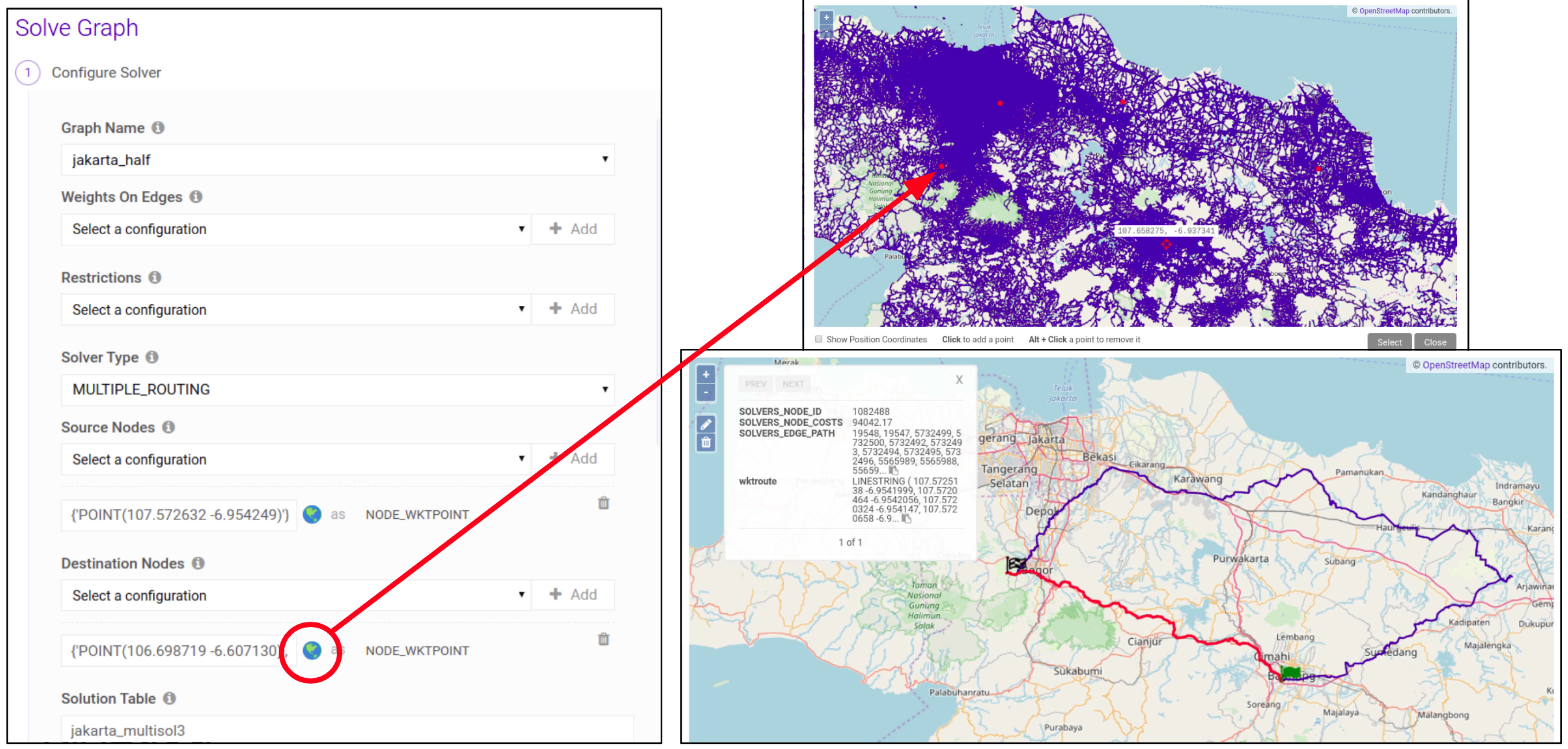}
    \caption{Graph requests can also be run via Kinetica-Graph-UI in which the identifiers are listed in the pull down with auto-completion for database schemas and graphical point picking capability over the graph network. (left) Solve-Graph UI widgets with ability to populate the multiple routing (travelling salesman) stop location by graphically picking over Jakarta road network. (right) The Jakarta road network graph and picked coordinates for Solve-Graph. (bottom) The result of Solve-Graph depicting optimal round-trip routing involving many stops across Jakarta.}
    \label{multiplerouting}
\end{figure*}

Kinetica-Graph is composed of just four generic and intuitive end-point schema APIs, namely, {\it Create, Solve, Query, and Match}. Each endpoint expects a particular set of components via identifier combinations annotated with a DB table schema or constant expressions. {\it Solve Graph} endpoint consists of low-level generic network solvers such as Dijkstra (shortest path), traveling salesman (with heuristics), back-haul routing, page rank, Markov chain probability, centrality between-ness, close-ness, inverse shortest path, all paths, Eulerian paths, etc. whereas the {\it Match Graph} uses more complex, specific purpose solvers that make use of combinations of generic solvers, such as Multiple Supply Demand Chain Optimization, Map Matching using Hidden Markov Chains, Origin-Destination time constrained routing, etc., as seen in Figure~\ref{solvers}. The end-point schemas are defined and designed in JSON format, and the implementation is done via Apache Avro serialization encoding into a C++ header representation. Various API forms of the endpoints are then created in R/C++/Java/JavaScript/Python language bindings~\cite{kinetica}. The bindings are usually wrapped by the endpoints' REST calls via HTTP requests as can be seen, e.g., in {\it Create Graph} endpoint in Figure~\ref{creategraph}. We have also provided the SQL equivalents of graph endpoints, in an intuitive manner as compliant as possible to the SQL standards in a Jupyter notebook like environment called Kinetica-Workbench, so that data analysts accustomed using SQL commands can also streamline graph calls into their ingestion, analysis, aggregation SQL pipelines with ease as shown in Figure~\ref{sqlcreategraph}. Graph requests can also be run via Kinetica-Graph-UI in which the identifiers are listed in the pull down with auto-completion for database schemas and graphical point picking over the graph network as depicted in Figure~\ref{multiplerouting}.

\begin{figure}
\centering
    \framebox{\includegraphics[height=0.2\textheight, keepaspectratio]{creategraph.pdf}}
    \caption{Kinetica-Graph Create/Graph Endpoint. Graph nodes are created from WKT point coordinates, and associated with labels, 'IN\_CAPITOL' or 'OUT\_CAPITOL' based on the result of the PostGres ST geometry distance function that measures the geodesic distance from the GPS location of the US Capitol, if the distance is less than 1km or more, respectively. Graph edges are created from the WKT-linestrings that match with the existing nodes within the 'merge\_tolerance' of 1 meter (0.00001 in degrees approx.)}
    \label{creategraph}
\end{figure}

\begin{figure}
\centering
    \framebox{\includegraphics[width=\linewidth, keepaspectratio]{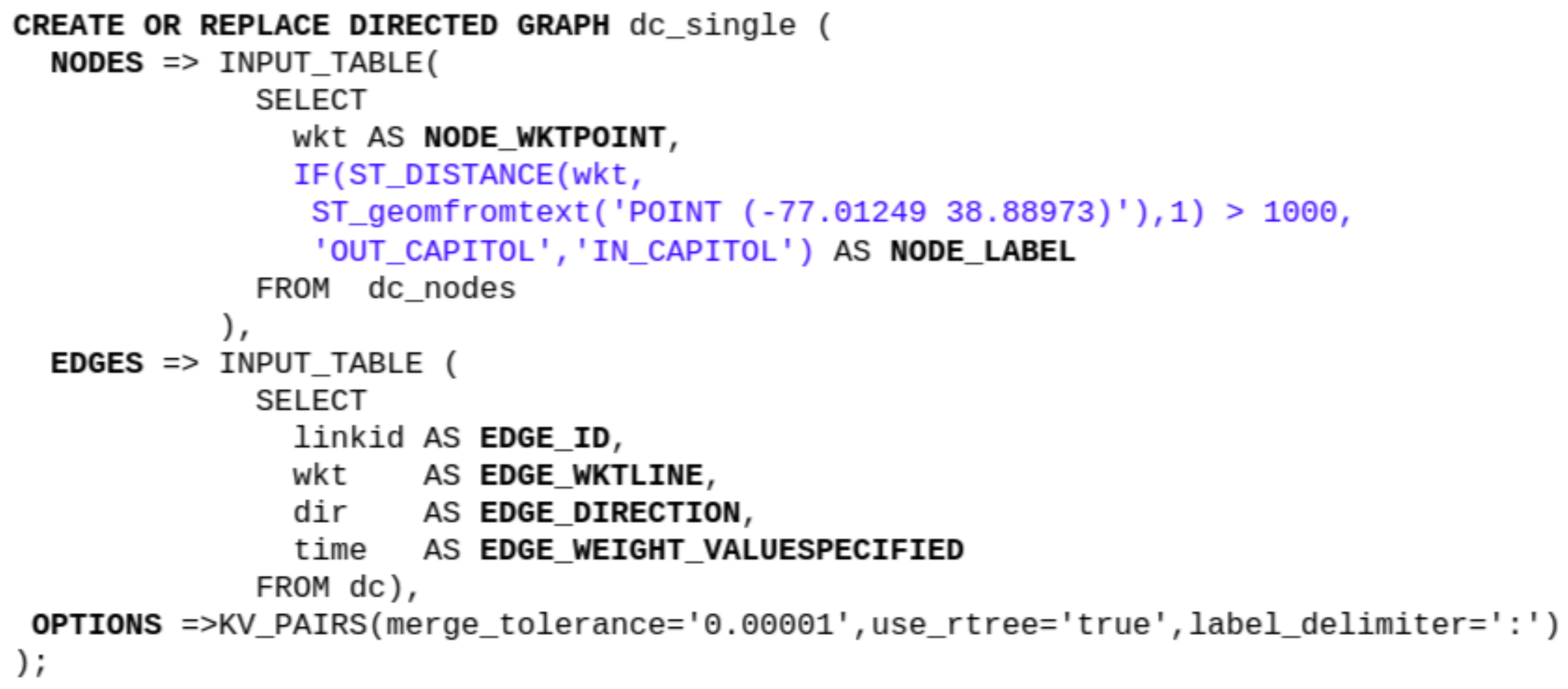}}
    \caption{Kinetica-Graph Create/Graph Endpoint in SQL syntax, equivalent of the RESTFUL request depicted in Figure~\ref{creategraph}.}
    \label{sqlcreategraph}
\end{figure}

\begin{figure*}
\centering
    \includegraphics[width=0.6\textheight, keepaspectratio]{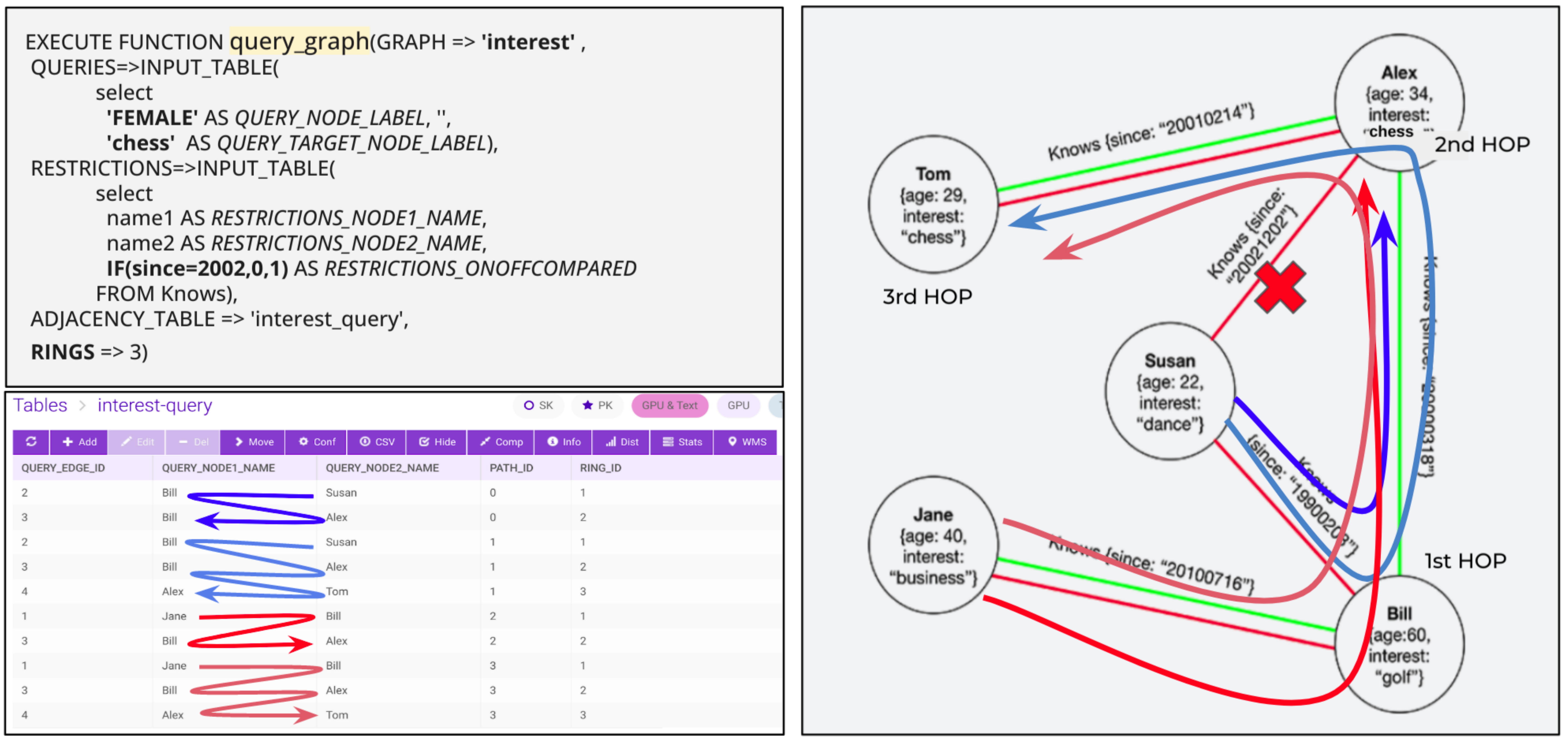}
    \caption{Query-Graph SQL endpoint with OLAP expression support for applying restrictions during many-to-many queries. Column \textit{Knows.since} that is not captured by the graph unlike columns \textit{Knows.name1}, and \textit{Knows.name2}, can still be accessed via Kinetica-DB in applying traversal restrictions: Many-to-many query shown in top left finds all the paths from `Susan' and `Jane' nodes that are labeled `FEMALE, to target nodes that are labeled `chess' within three hops. Restrictions applied on the edges whose `Person' nodes knows each other since $2002$ with the help of `IF' OLAP function; four paths are found by skipping the restricted edge shown in red cross on right, two from `Susan' and two from `Jane' shown in the table interest\_query in blue and orange paths, respectively (bottom left).}
    \label{manyqueries}
\end{figure*}

Kinetica-DB is also able to run many to many queries at-scale. Query syntax is different but functionally compliant to cypher queries, in that one can have query-time restrictions akin to the `WHERE' clause very easily and flexible in Kinetica-DB as shown in Figure~\ref{manyqueries} where multiple paths are found from `FEMALE' nodes to persons whose interests are in playing `chess' (node lables attached at the time of Create-Graph) while restricting on edges with the help of the OLAP functions provided by Kinetica-DB. 

\section{Server-Client-Server architecture}
\label{Section:communication}

\begin{figure*}
\centering
    \includegraphics[width=0.7\linewidth, keepaspectratio]{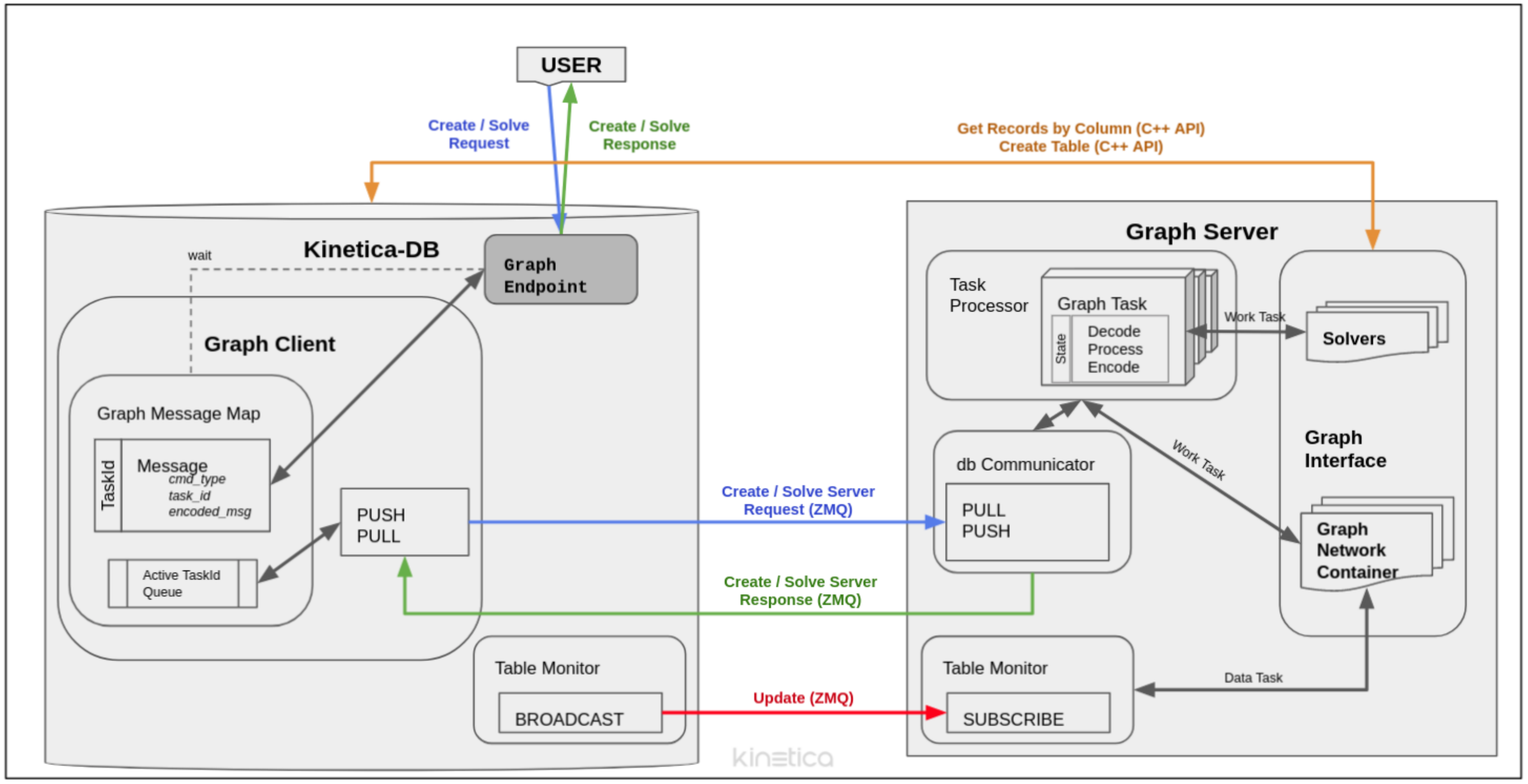}
    \caption{Client-Server hybrid architecture of Kinetica-Graph with Kinetica-DB using ZMQ and C++ Apis from Kinetica-DB}
    \label{architecture}
\end{figure*}

Kinetica-Graph Server(s) are implemented to communicate with the Kinetica-DB via its C++ APIs using pull/push pattern of ZMQ socket communication~\cite{zmq} for expression and I/O support of the database as depicted in Figure~\ref{architecture}. Graph tasks are encoded into byte stream using Avro~\cite{avro} serialization. Graph-Client is the interface inside Core-DB pushing user requests to the server by adding extra server side parameters. Graph-Servers pull from the socket channel and decode the messages back to their original specific tasks in a threaded do-run loop so that the incoming tasks had been processed based on their priority order and never get lost. Graph-Servers also provide the proper locking mechanism on the graph network objects via Graph-Interface. The Task-Processor of each Graph-Server runs the concurrent tasks in their own queue by waiting on each other properly (See Figure~\ref{architecture}).

This communication style ensures the tasks to be processed in the queue in a non-blocking fashion. For example, if a Solve-Graph is requested while the Modify-Graph is updating the graph-edge topology, solve has to wait till the graph finishes updating due to the locks on the graph entities. However, as soon as the scoped graph locks are lifted, many solves on the same network can run concurrently as they operate on the same constant graph object in different threads.

\begin{figure}
\centering
    \framebox{\includegraphics[width=\linewidth, keepaspectratio]{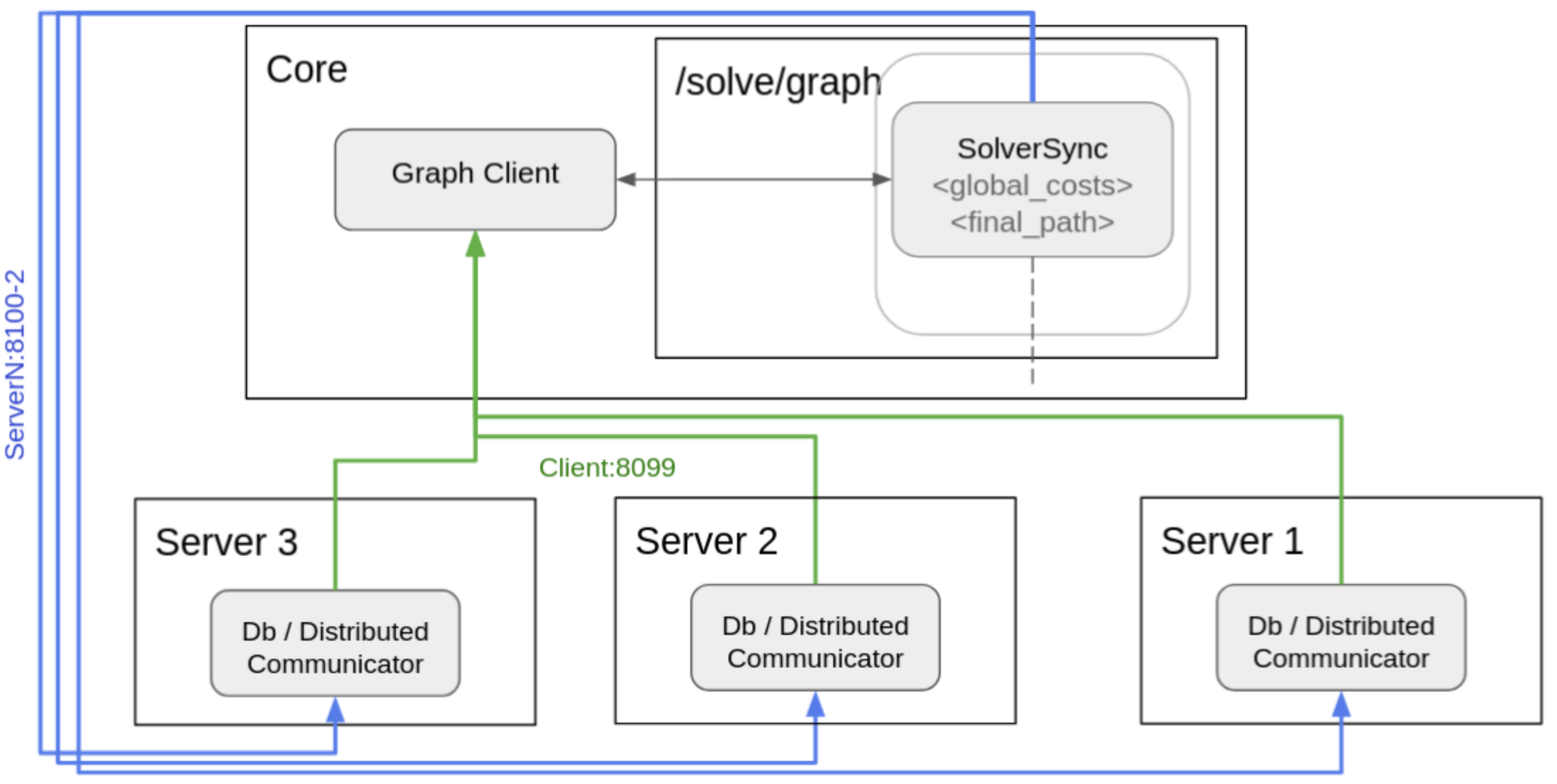}}
    \caption{Server-Client-Server ZMQ communication for distributed solves}
    \label{distribcomm}
\end{figure}

Similar but more complex pattern of socket communication architecture is devised for distributed solves, as shown in Figure~\ref{distribcomm} where the Graph-Client orchestrates communication from/to many Graph-Servers that message over dedicated read/write ports in a sequential loop. This type of many-to-one-to-many communication pattern ensures the abort-able conditions to be observed in a coherent fashion in which the conditions would not be raced among the  Graph-Servers. 

A typical distributed solve relies on the fact that the costs at the duplicated nodes (will be discussed in Section~\ref{Section:rebalancing} and~\ref{Section:distributed}) will converge after iterations on many servers. For example, if there is an update on the cost of a duplicated node reported by one of the servers, that server sends the message to the Graph-Client, and the Graph-Client sends it to the port that every other server reads from and depending on the closure of the duplicated node, the cost value could be used as the trigger (new front) on that server's own solve, and so and so forth. This process of servers updating themselves continually repeats until convergence is reached. We will cover the algorithm of the distributed solves in more detail in Section~\ref{Section:distributed}.

\section{Distributed graphs and re-balancing algorithm}
\label{Section:rebalancing}

\subsection{Replicated and Partitioned Graphs}
\begin{figure}
\centering
    \includegraphics[width=0.7\linewidth, keepaspectratio]{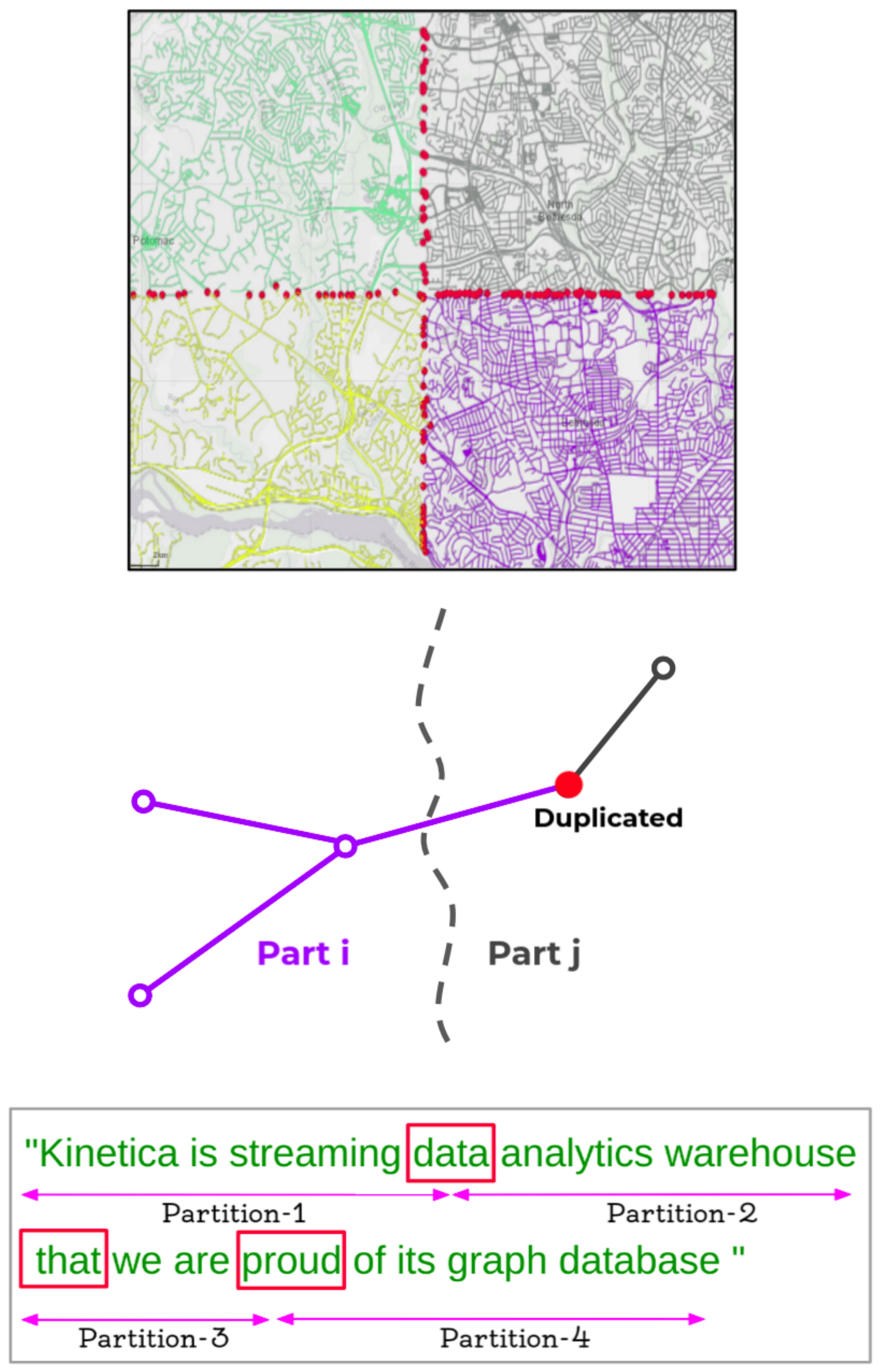}
    \caption{Internal (implicit) partitioning algorithm ensures that every edge belongs to only one partition: DC metropolitan area partitioned into four servers (top), blue edge belongs to Part $i$ since $i<j$, and the red node is set as a duplicated node by each graph server running server side Create-Graph requests concurrently (bottom), property graph constructed using external (explicit) partitioning by  edges linking words;  duplicated nodes as 'data', 'that' and 'proud' shown within red rectangles (bottom).}
    \label{duplicated}
\end{figure}

\begin{figure}
\centering
    \includegraphics[height=0.4\textheight, width=\linewidth, keepaspectratio]{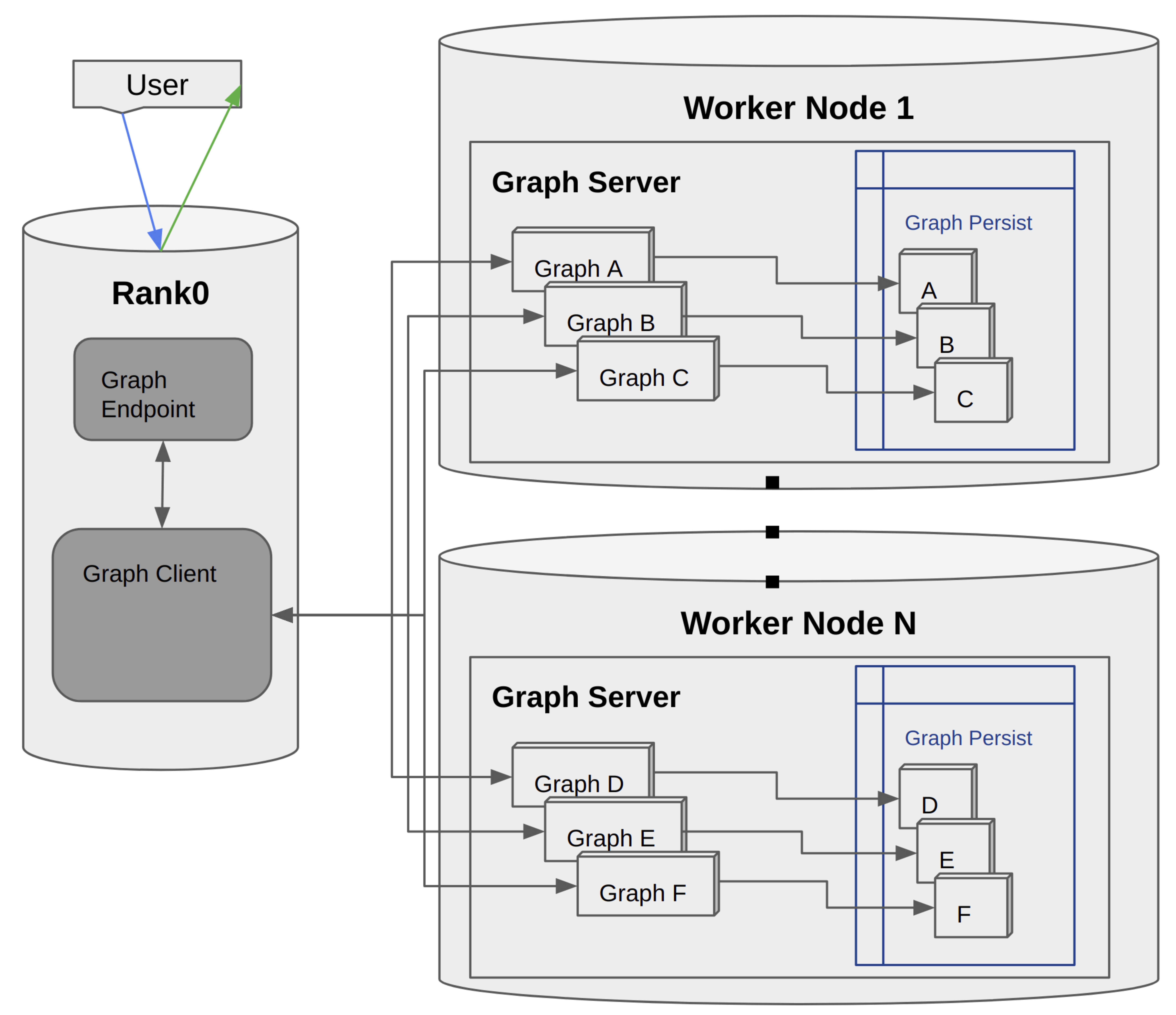}
    \caption{Distributed Graphs - Two flavors exist: replicated or partitioned. Replicated graphs are identical copies of the same graph in each server and the partitioned graphs are sub-graphs in each server linked via the duplicated nodes.}
    \label{distributedgraphs}
\end{figure}

Graphs can be replicated or partitioned among many graph servers. Each graph server is a separate executable and can be instructed to be installed on a rank (node) where its persist (serialized byte dump + metadata) can also reside separately as shown in Figure~\ref{distributedgraphs}. Replicated graphs are identical copies of the same graph in each server, and any solve or query input is split among them to reduce the input size per server, so that the speed of computations could be increased at-scale. However, when the graph size is of concern, partitioning the graph into sub-graphs is the only other alternative. We have devised a topology partitioning data model such that every graph edge can belong to only one server, and topology connections is continued across servers via nodes that are duplicated at the inter-server junctions as shown in Figure~\ref{duplicated}. 

\subsection{Internal and External Partitioning}

\begin{figure*}
\centering
    \includegraphics[width=\linewidth, keepaspectratio]{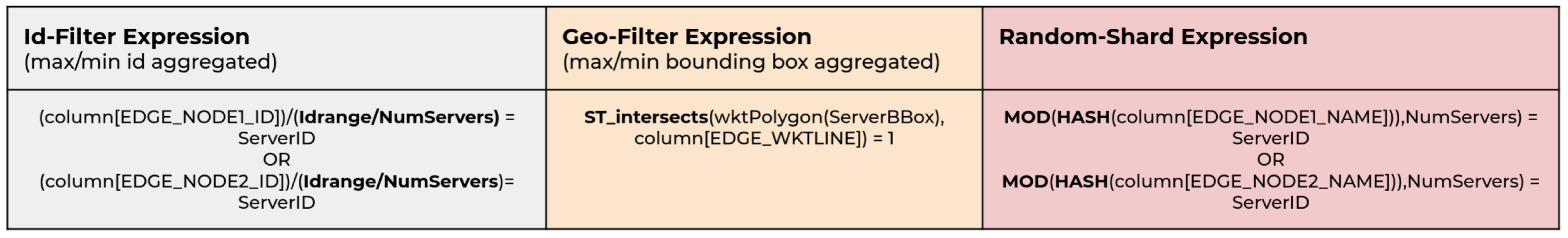}
    \caption{Various distributed internal partitioning filters; the primary edge table(s) is processed to find the global aggregated min/max ids (Id-Filter) or min/max wkt bounding box (Geo-Filter) parameters followed by the distributed OLAP filtering based on the graph type spawned at each graph server. Sub-graph edges in each partition is then created from these views. Another internal partitioning scheme (Random-shard) is the random distribution on string based social graph edge nodes, in which OLAP based MOD and HASH functions are employed in the filter expression for creating the respective views for each graph server.}
    \label{filters}
\end{figure*}

There are both internal (implicit) and external (explicit) partitioning schemes available in Kinetica-Graph. Internal partitioning can be done in three different flavors depending on the type of the graph, that can be deduced by the edge identifiers in the Create-Graph request call, namely, id range, geometric bounding box range or random sharding as shown in Figure~\ref{filters}. There is no guarantee, however, that the partitions created in any of these three ways would have the least number of duplicated nodes, as it is one good indicator for the effectiveness of the partition. We can speculate that the best partitions are defined to have minimum inter-communications to carry out a distributed task with least number of iterations, hence total number of duplicated nodes over the size of the graph is considered to be a good score for efficient partitions. Any arbitrary partition scheme by the user can easily be applied by imposing explicit partitioning schemes using specific identifiers when creating the graph such as \textit{EDGE\_PARTITION},  and \textit{NODE\_PARTITION\_BOUNDARY}. For partitioned graphs, Create-Graph response is a set of partitioned graphs (same name) on each server and the partitioning score, reported as the aggregated sum of the duplicated nodes.

\begin{figure}
\centering
    \framebox{\includegraphics[width=0.7\linewidth, keepaspectratio]{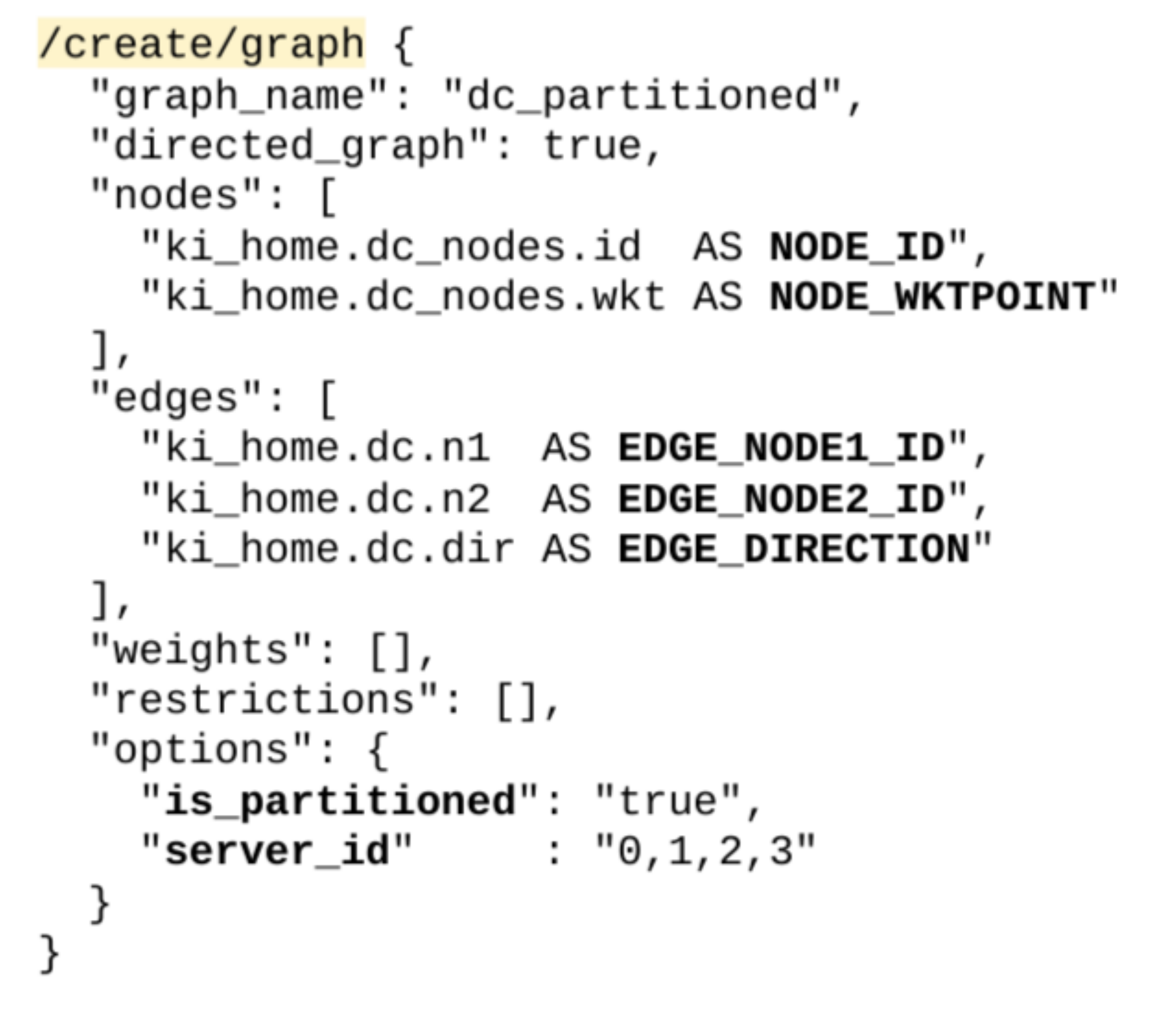}}
    \caption{Create-Graph request invoking Id-range internal scheme for partitioned graphs over four servers specified in the options.}
    \label{createpartgraph}
\end{figure}

The clustering of the partitions has a huge impact on the speed of the distributed solves and queries since scalar values of the analytics has to be transferred across adjacent partitions many times for convergence. If the solve or query graph traversals need to jump back and forth across partitions frequently due to inter-mingling of the partitions with poor clustering characteristics, it would take many more iterations to converge as seen in Figure~\ref{rebalanced}~(b) that shows the DC metropolitan area road network graph created using node ids by the Create-Graph request depicted in Figure~\ref{createpartgraph}. In order to visually inspect the partitions, we have purposefully associated the node ids with spatial coordinates, but otherwise, the graph is merely constructed from integer pairs of node ids. Each edge in the picture is colored based on the partition that is calculated by the id-range evenly distributed  over the number of servers. Hence, this type of partitioning results in highly interlaced partitions, with very poor clustering that crucially needs a balancing algorithm to improve  clustering which will be described in the next~Section~\ref{subsections:balancing}.

\begin{figure*}
\centering
    \includegraphics[height=0.5\textheight, width=\linewidth, keepaspectratio]{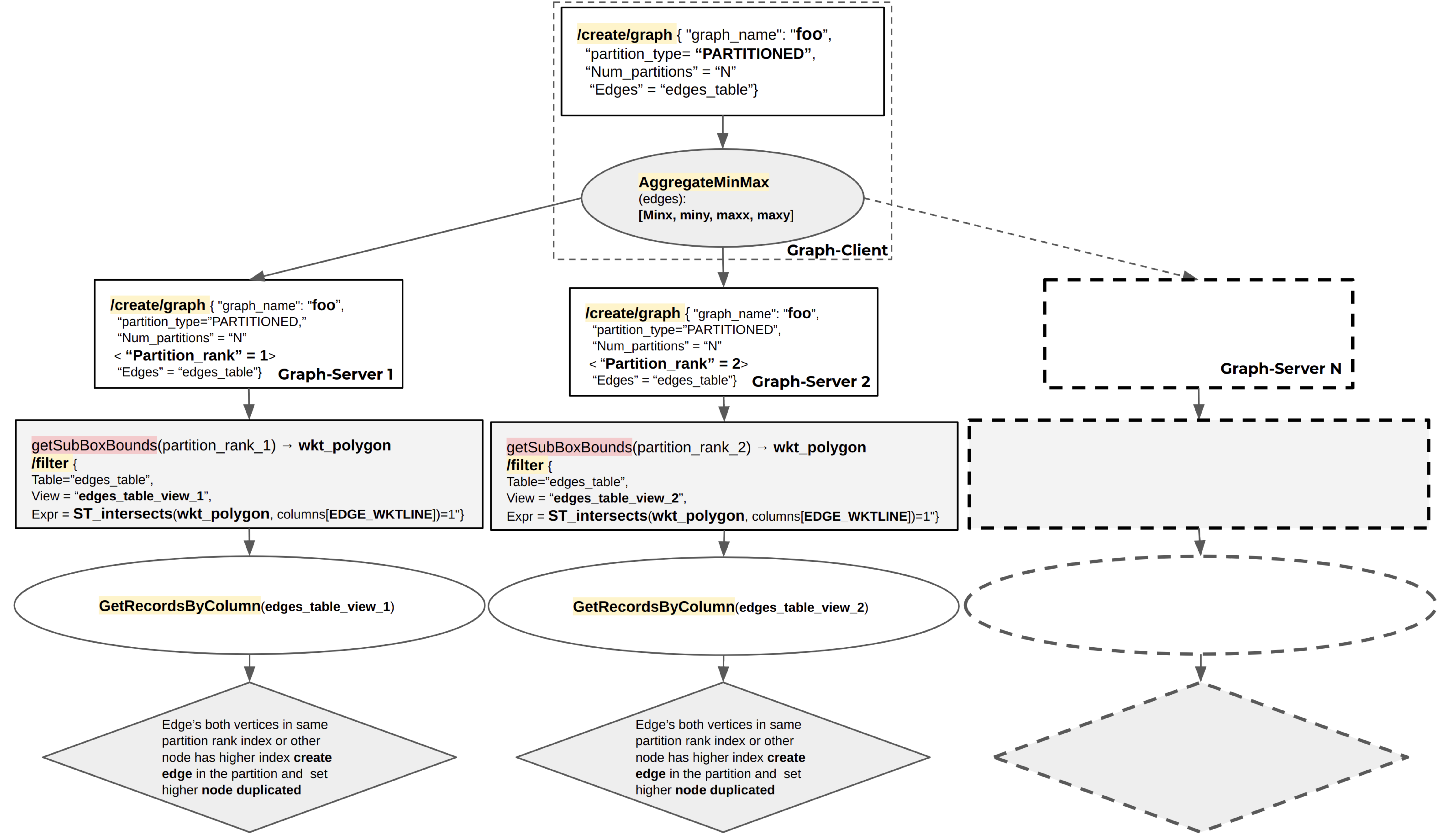}
    \caption{Complete Partitioning algorithm using GEO filters: Graph-Client receives the partitioned Create-Graph request and determine that the edges are created from WKT geometry columns and employs distributed AggregateMinMax OLAP call to extract the global bounding box coordinates. Server side Create-Graph calls are rectified and sent to each Graph-Server so that they themselves can run distributed OLAP filtering to create views within their sub bounding boxes via the ST\_intersect geometry function. Edges are created from these views by further analyzing if both nodes of an edge fall within the bounding box or otherwise, the outside node is set to be a duplicated node if it has a higher rank (server-id) that that  of the partition.}
    \label{wktpartitioning}
\end{figure*}

Before moving on to the balancing algorithm, it is worth noting how a slightly more efficient geometric bounding box partitioning algorithm is envisaged as one of the other implicit and distributed partitioning heuristics summarized in Figure~\ref{wktpartitioning}. Basically, the geometric extent of the graph is computed using the distributed OLAP function, namely, \textit{AggregateMinMaxGeometry} and this bounding box is further divided into $n~\times~m$ lattices along $x$ and $y$ among multiple graph servers in an ad-hoc manner. Each graph server then executes a set of  internal Create-Graph calls by running over the view generated by another distributed OLAP filtering whose expression consists of the PostGres \textit{ST\_intersects} function to peel off the corresponding quadrant from the input. However, this filtering expression is still not enough to create the final \textit{non-overlapping} partitioning since the geo-spatial polygon intersection of the graph network by a box results in edges that are straddling the lattice boundaries of adjacent servers. Therefore, an additional step is required to ensure that every graph edge belongs to only one partition (server) in a consistent manner comparing the lexicographical order of the server ids of its nodes as depicted in Figure~\ref{wktpartitioning}. 

\subsection{Balancing partitions}
\label{subsections:balancing}

The aim of balancing the partitions is to divide graph equally with the least number of duplicated nodes along inter-server boundaries. This is a non-deterministic optimization problem and becomes even harder when the partitioning should be done in a distributed manner without bottle-necks. We have devised a clustering algorithm ensuring equal division with 'reasonably' minimal total number of duplicated nodes using the results of a distributed shortest path solver over the unbalanced partitions. Here are the steps of the balancing algorithm:

\begin{itemize}
\item[-] Step 1. Distributed shortest path solve without path aggregation from a chosen source (default source node can be overridden) to all the nodes in the graph as targets over the unbalanced partitioned graphs.
\item[-] Step 2. Write out the graph nodes with solved cost values at each record into a \textit{nodes} table, and graph edges with node ids, as \textit{edges} table. Ensure all the node and edge labels along with other graph attributes are preserved in these tables as additional columns.
\item[-] Step 3. Renumber node ids of the \textit{edges} table using ascending sort (distributed merge sort) of the cost from the \textit{nodes} table.
\item[-] Step 4. Recreate partitioned graph from the \textit{edge node identifiers} that use the node ids computed in previous stage and use id range implicit partitioning to generate the balanced partitions. Create-Graph endpoint using the 'recreate' option automatically swaps unbalanced graphs, with the balanced partitions.  
\end{itemize}

\begin{figure*}
\centering
    \includegraphics[width=\linewidth, keepaspectratio]{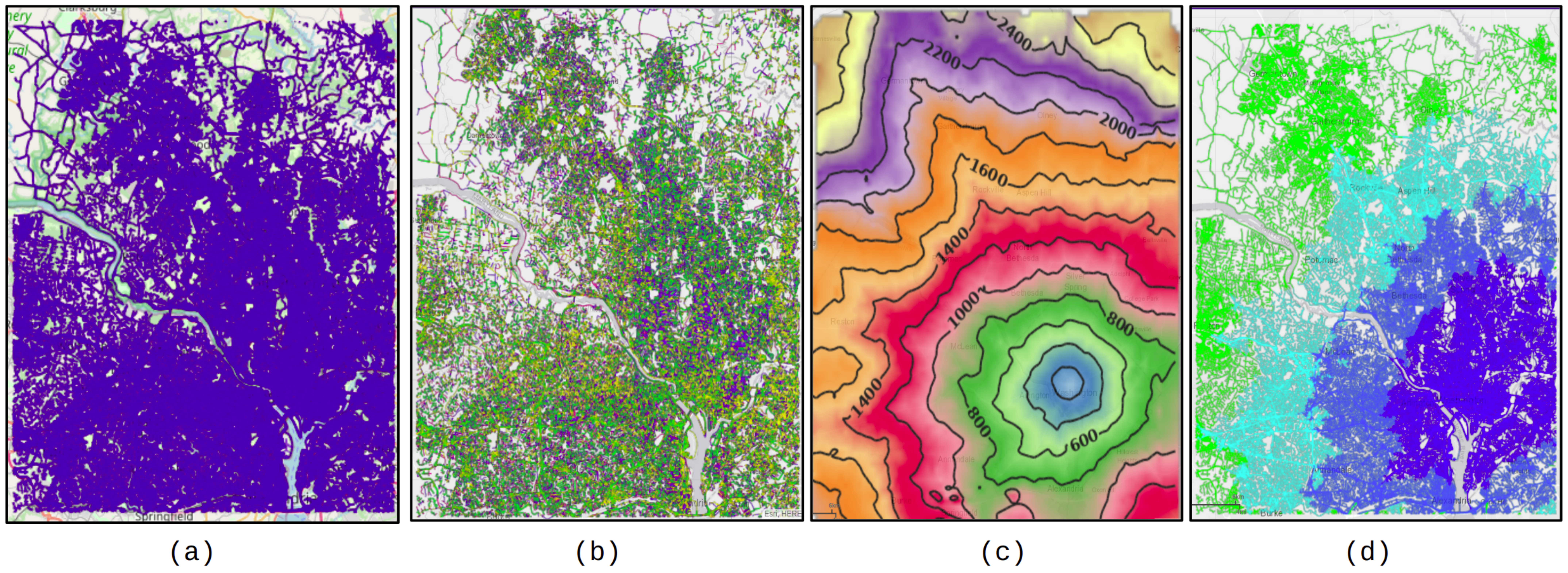}
    \caption{Balancing the partitions: (left) graph topology, unbalanced partitions by id-range internal partitioning (colors indicate partitions), contours of the sssp solve results over unbalanced partitions, balancing using the iso-levels of the solve as a separator such that each partition would have equal number of edges (right). The clustering of the partitions in colors can easily be seen to have improved between the second image from left (unbalanced partitions) to the image on the far right (balanced partitions).}
    \label{rebalanced}
\end{figure*}

The results of clustering after balancing can be seen on the color codes of the edges in the partitions in Figure~\ref{rebalanced}.(d). The improvement is markedly visible when compared to the interlaced colors (partitions) in the unbalanced graphs shown in Figure~\ref{rebalanced}.(b) where clustering is very poor for the DC road network graph. Node ids are renumbered based on the ascending cost values from the results of the distributed shortest path solve (Dijkstra)  depicted in Figure~\ref{rebalanced}.(c). 

The impact of balancing is huge on the speed of the distributed solves and queries. The comparison numbers will be given at the end of Section~\ref{Section:distributed} after the explanation of the distributed solver algorithm for completeness. Though, based on our findings that we gathered from our extensive testing, we can confidently state that the difference is almost two orders of magnitude in the solution speed, i.e., the solver over balanced partitions runs approximately  $100$ times faster than the unbalanced (random) partitions.

\section{Distributed solver}
\label{Section:distributed}

We have opted a priority queue implementation for our version of the Dijkstra solver which seems to supersede parallel queue implementations~\citep{mapmatching, felner}. In general, the Dijkstra Condition (DC) on each node $v_i$ can be specified by Equation~(\ref{dijkstra_eqn}) which states that the cost $d_i$ can not be greater than the minimum of the cost of any incoming nodes connected to $v_i$ via the edge's weight $w_{ij}$. The DC condition is satisfied in a breadth first search manner by the Dijkstra-$\mathscr{D}$ kernel originated from the source (start) node and terminated at the destination (end) node. The modification of DC for the distributed graph case, is simply the update of the cost values $d_i$ at the duplicated nodes among adjacent Graph-Servers $G_k$ (partitions) via an iterative process as depicted in Equation~\ref{distribuited_dijkstra_eqn}. 

\begin{align}
d_{i} = (d_j + w_{ij})~ \mid ~w_{ij}\colon v_j \mapsto v_i,   \in N(v_{i}) \nonumber \\
\mathscr{D}_{start,end} =  \min_{v_i \in G(V,E) \mid_{start}^{end}} \Big(d_i\Big)  \label{dijkstra_eqn}
\end{align}

\begin{align}
d_{i}^{new}\mid_{G_j} = \min (d_i\mid_{G_j}, d_i\mid_{G_k}) \colon d_i \in G_k \cap G_j  \label{distribuited_dijkstra_eqn}
\end{align}

\begin{figure*}
\centering
    \includegraphics[height=0.45\textheight, keepaspectratio]{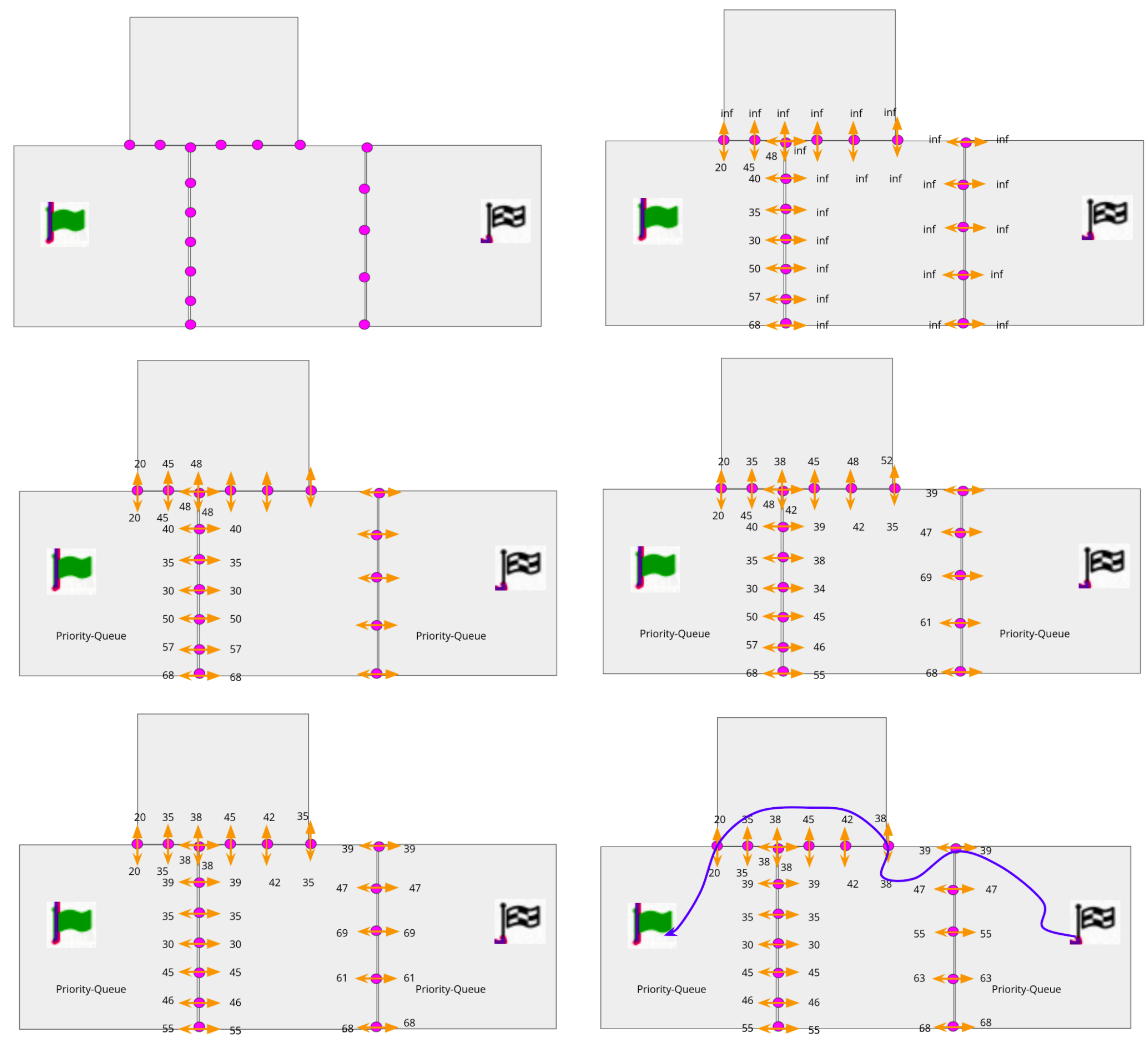}
    \caption{Steps of a priority queue based distributed shortest path solver algorithm: (top left) Four partitions with duplicated nodes depicted in red, (top right) Only the left most partition's Dijkstra solver works triggered from the source being at zero and all other nodal cost values at infinity, (mid left) The nodal cost values gets transferred to the adjacent partitions at the duplicated nodes if a lesser value is found, (mid right) priority queue solver of the adjacent partition gets initiated from the updated node and cost pair set, (bottom left) Cost values at the duplicated nodes gets exchanged whichever value is smaller from adjacent partitions triggering yet another solve wherever the update happens, (bottom right) There is no more updates needed which means convergence is reached, the path is then aggregated starting from the target and revisiting partitions until the source is hit depicted as the blue line.}
    \label{sssp}
\end{figure*}

First, the partition that contains the source node is located to start the process. The solver of the partition receives the front pair $(d_i, 0)$ and populates its heap structure with the front and solves towards all the other nodes in the partition as depicted in Figure~\ref{sssp}.(top). The rest of the unvisited nodes in the partitions have the cost value at infinity. Other Graph-Servers are waiting to check and update if any of their duplicated nodes to have a lower cost from the adjacent partitions, concurrently. After the initial solve is finished where the source node is contained, the costs can spread to replace the infinite costs and thereby trigger the solves at adjacent partitions. Updated costs at the duplicated nodes are paired to populate the new fronts of the adjacent solvers as shown in Figure~\ref{sssp}.(mid). The process of concurrent runs at each Graph-Server continues until no more cost updates are found which means that the Dijkstra condition is satisfied globally across all partitions. Finally, shortest paths found by aggregating that starts from the target back to the source and stitching through the duplicated nodes using the result of the Dijkstra solver at each Graph-Server. This back tracking process of the path aggregation algorithm stops when it reaches the source node as depicted in Figure~\ref{rebalanced}.(bottom). 

\begin{figure}
\centering
    \includegraphics[height=0.3\textheight, width=\linewidth, keepaspectratio]{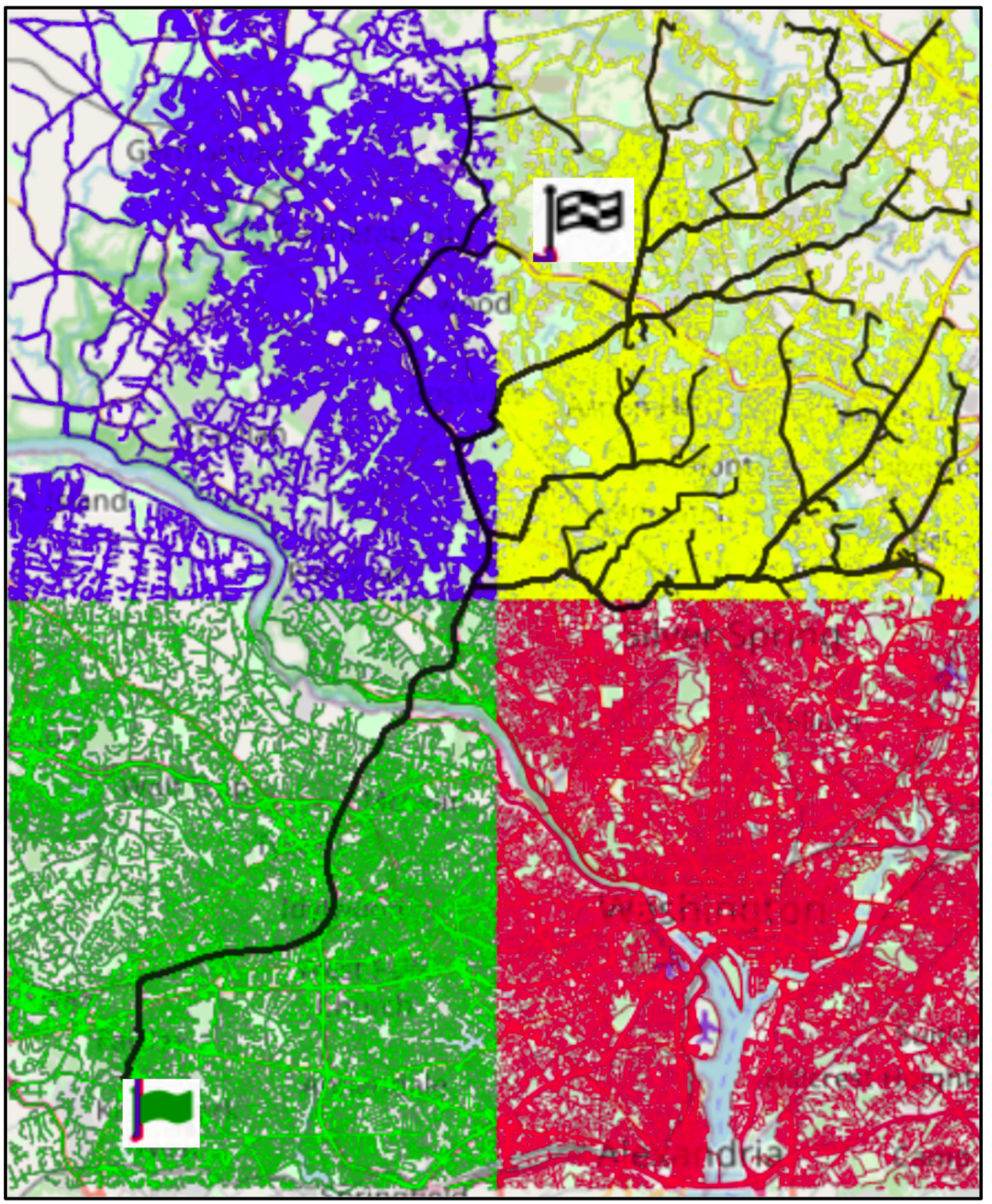}
    \caption{The shortest paths from single source to all targets over four partitions. Paths may go in and out of partitions. The shown paths in black are just the first 100 destinations in the third quadrant (partition).}
    \label{manytargets}
\end{figure}

\begin{figure}
\centering
    \includegraphics[height=0.4\textheight, width=\linewidth, keepaspectratio]{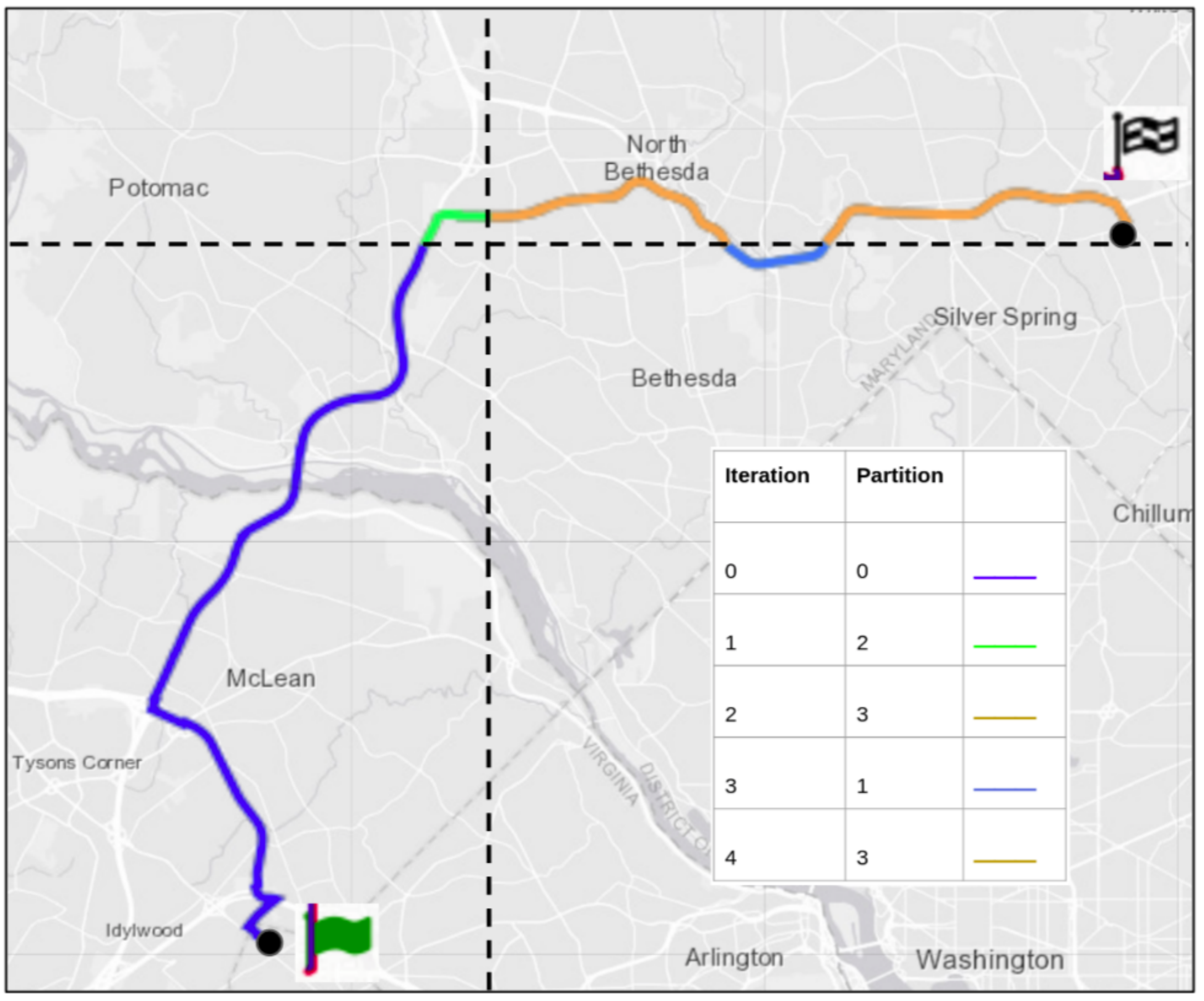}
    \caption{Aggregating the shortest path by back tracking; partition $3$ is visited twice from the source located at the lower left partition $0$}
    \label{pathaggregation}
\end{figure}

\begin{figure*}
\centering
    \includegraphics[width=0.9\linewidth, keepaspectratio]{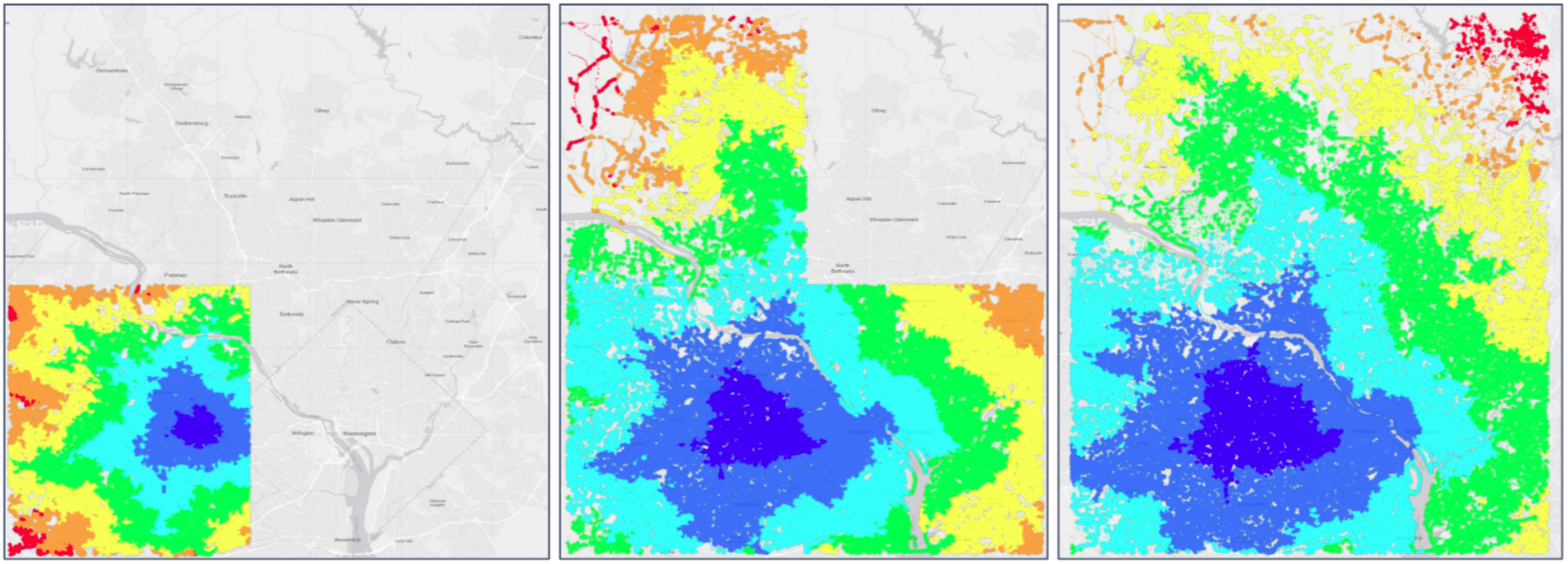}
    \caption{The propagation of the cost in the first three iterations across partitions in color codes from left to right; each iteration of the distributed solve updates and advances the cost field to a more converged state. Note that the distant nodes move away further in red color from left to right as the iterations advance.}
    \label{propagation}
\end{figure*}

\begin{figure}
\centering
    \includegraphics[height=0.2\textheight, width=\linewidth, keepaspectratio]{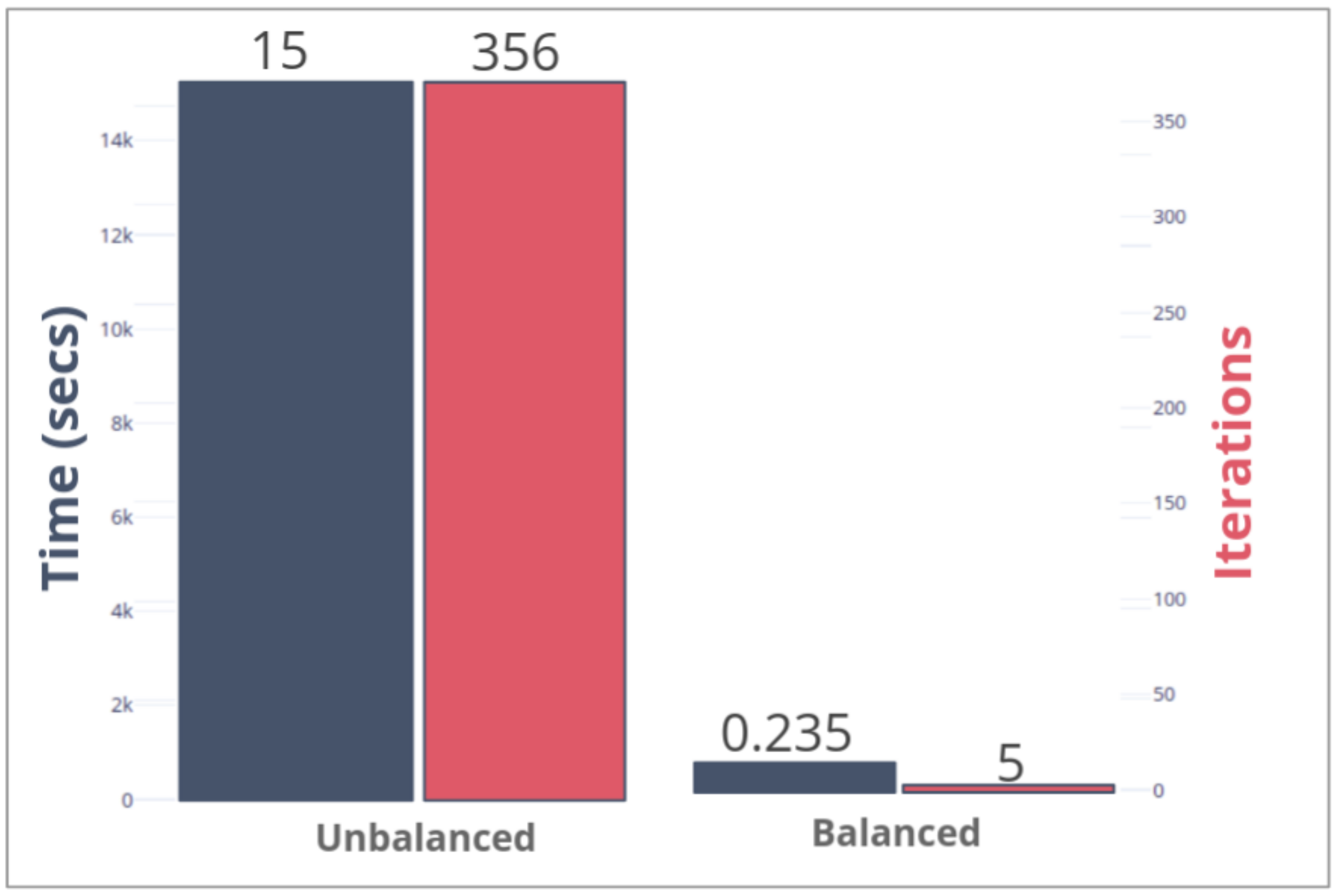}
    \caption{SSSP run to all targets with path aggregation for balanced vs unbalanced graph comparison on the total number of iterations and time till convergence using four Graph-servers.}
    \label{timeiter}
\end{figure}

The paths from the same source to many targets can also be found as shown in Figure~\ref{manytargets} in which the metropolitan area of DC road network is divided by four Graph-Servers using the bounding box partitioning scheme, and first hundred targets in the upper right quadrant is chosen as the target nodes. The shortest paths emanating from the source located in lower left to all the targets in the upper right partition can also be seen in Figure~\ref{manytargets}. One of these paths can be investigated to see how the same server is visited more than once in aggregating the path over the scalar cost field where Dijkstra condition is satisfied across the partitions in Figure~\ref{pathaggregation}. The propagation of this scalar cost field during concurrent iterations over the Graph-Servers can be seen in Figure~\ref{propagation} as the red colored nodes, signifying the highest distance (or time) cost move away as the cost gets corrected and spreads to distant quadrants (partitions). The effect of the balancing algorithm explained in the previous Section~\ref{Section:rebalancing} versus the unbalanced random sharding on the speed of the solve is markedly different. If the graph is created from the node ids (id range) as depicted in the Create-Graph call of Figure~\ref{createpartgraph}, versus balancing partitions by Repartition-Graph call whose steps are summarized in Section~\ref{Section:rebalancing}, the difference in solver performance is two orders of magnitude both in the total number of iterations and total time to convergence as tabulated in Figure~\ref{timeiter}. In the geo-graphs particularly, the performance difference between balanced versus bounding box partition is not that much, however, balanced solve is still twice as fast if not hundred times as compared to the cases of id-range or random partitioning in social graphs.  

\begin{figure}
\centering
    \includegraphics[width=\linewidth, keepaspectratio]{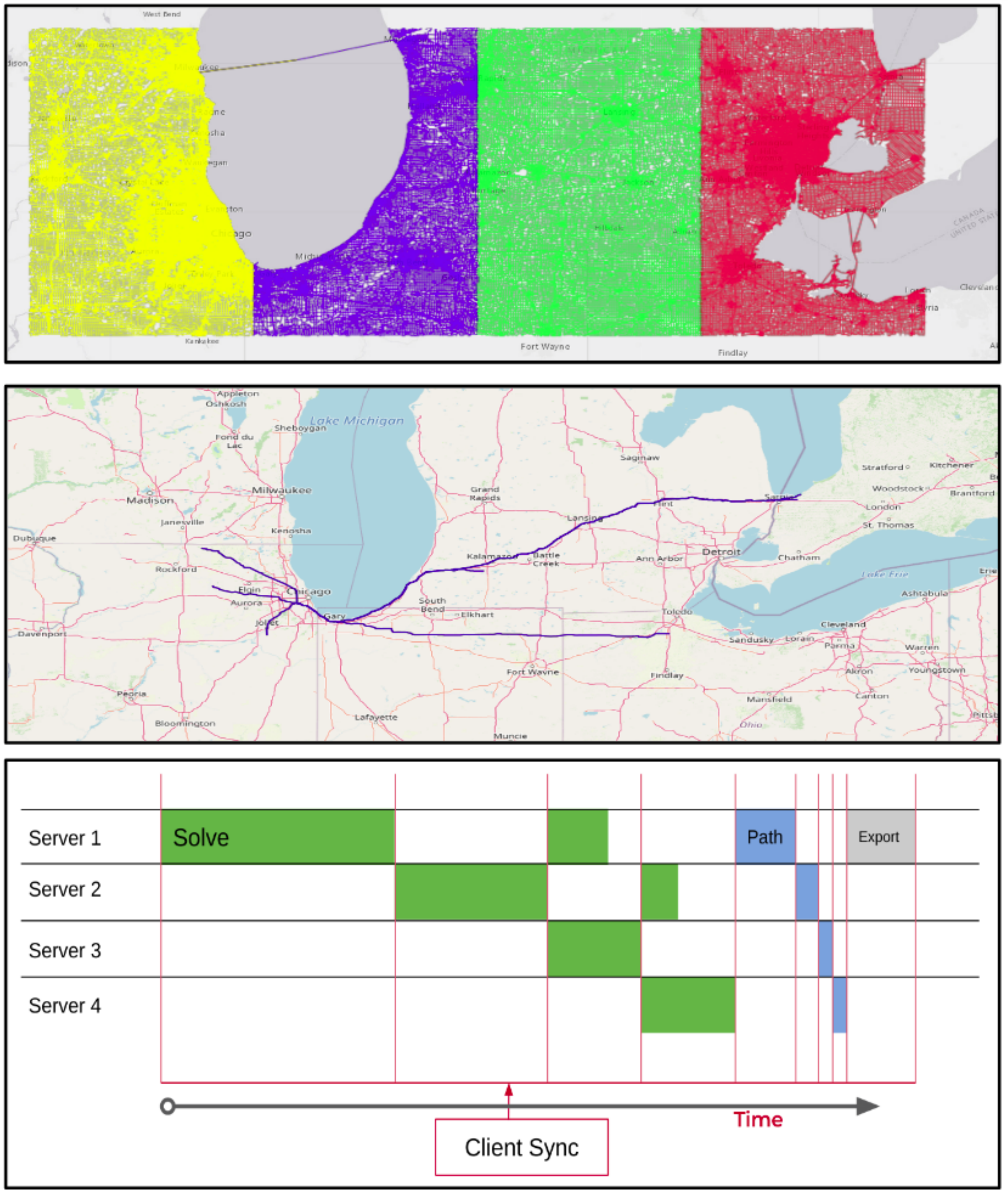}
    \caption{Distributed shortest path solves across Great Lakes with 4 servers using bounding box range subdivision (top), The shortest paths from one source to five destinations (mid), the distributed solver's tracer in time analysis, where in each cycle there is either one or more servers concurrently run and exchange cost values across duplicated nodes and path aggregation using the back tracing starting from the targets  (bottom).}
    \label{distribsolve}
\end{figure}

\begin{figure}
\centering
    \includegraphics[width=\linewidth, keepaspectratio]{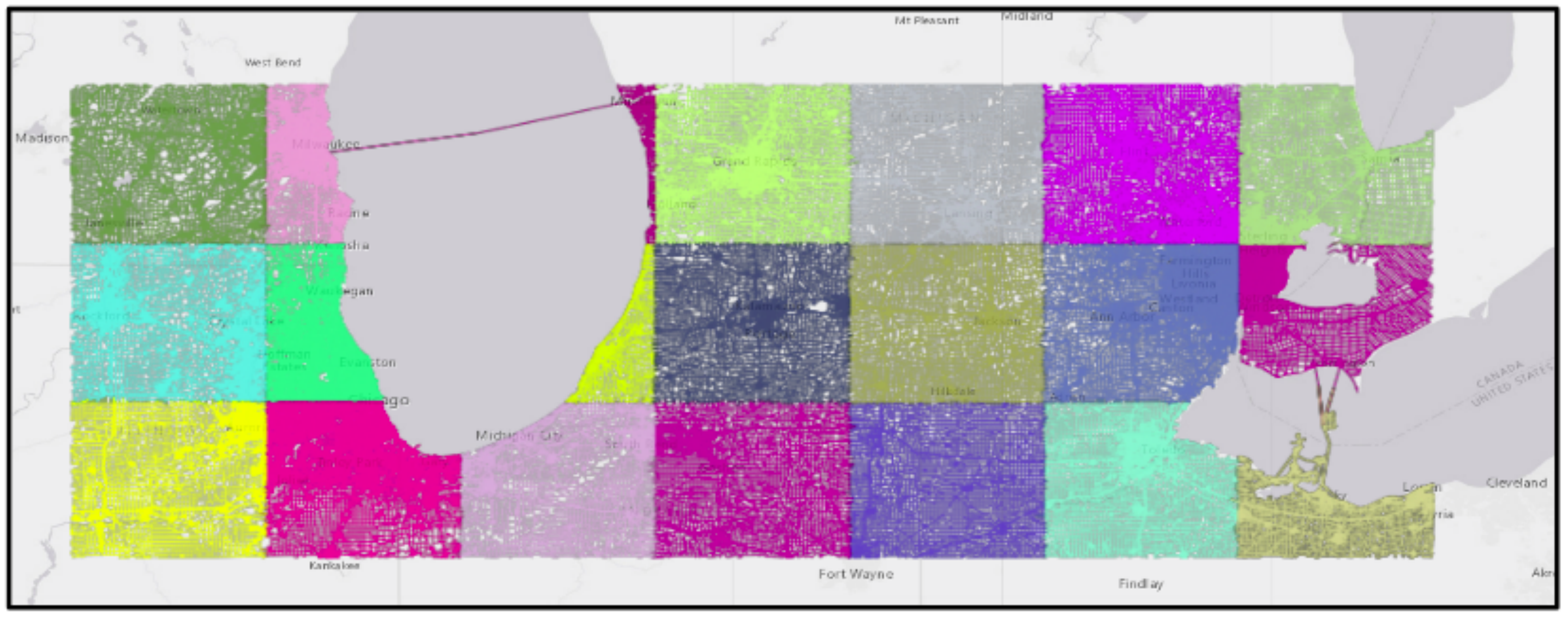}
    \caption{Great lakes area can be divided into more partitions to speed up the distributed solves based on the availability of the resources. Graph is partitioned based on the bounding box division into $3\times7$ partitions.}
    \label{part3x7}
\end{figure}

Another example of running distributed shortest path solver over the four partitions of the great lakes area is shown in Figure~\ref{distribsolve}. The time tracer analysis across four servers depicts how servers run concurrently while updating cost values across the duplicated nodes. Note that the path aggregation process is sequential and may not be insignificant compared to the total solve time. Distributed solves can be sped up using more partitions as shown in Figure~\ref{part3x7} based on the availability of the resources.

\section{Results and Conclusions}

The success of Kinetica-Graph is built on top of its fixed memory topology data-structure framework that has no memory degradation in dynamic graph updates as discussed in Section~\ref{Section:topology}. We also have adopted a novel network agnostic graph grammar and wrapped it with SQL syntax functionally compliant to Cypher queries, discussed  in detail in Section~\ref{Section:hybrid}. The integration of graph operations with the OLAP engine using the SQL syntax is our unified solution and a game changer. 

\begin{figure}
\centering
    \includegraphics[width=\linewidth, keepaspectratio]{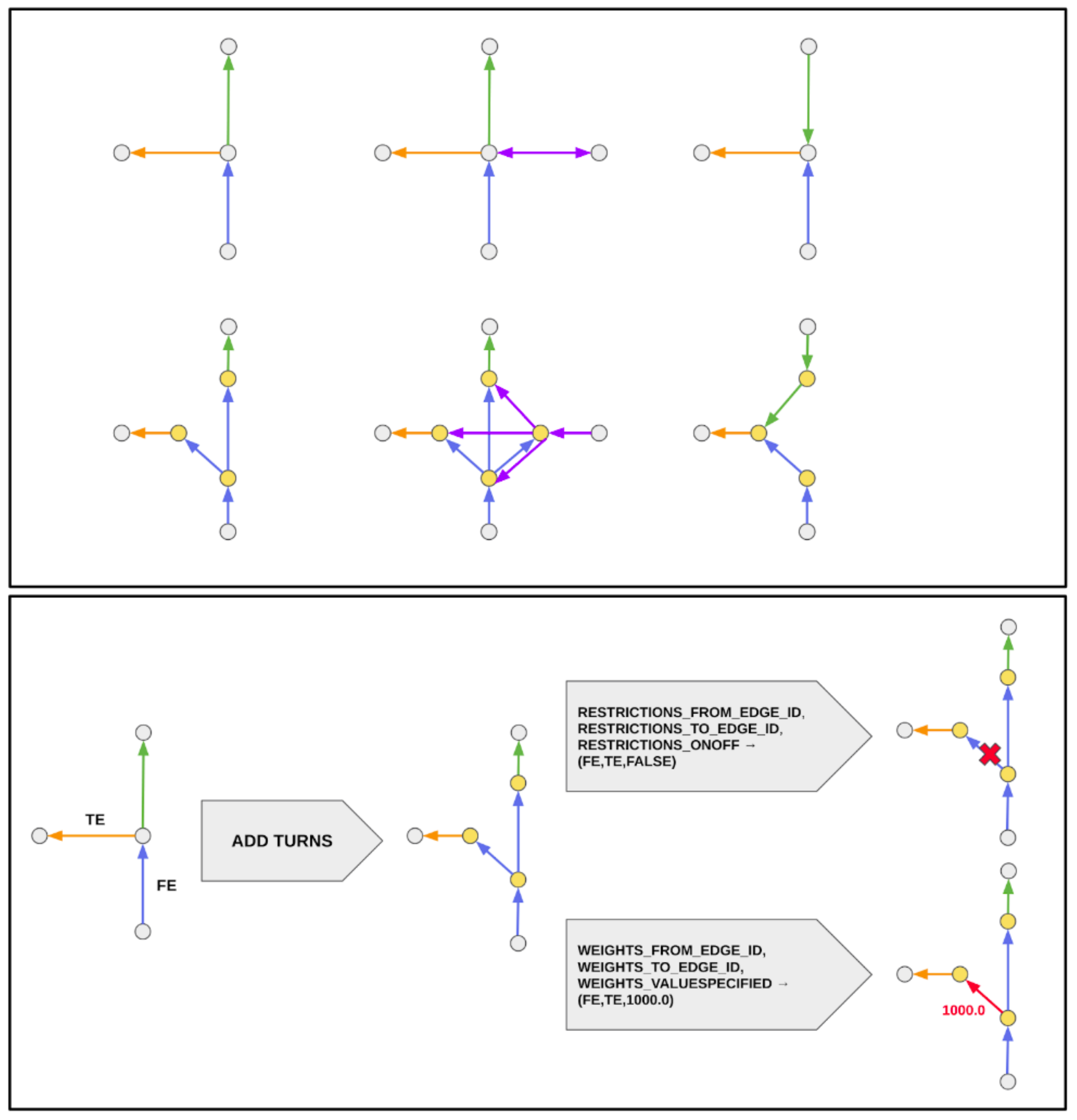}
    \caption{Adding automatic turn restrictions via dummy nodes/edges (top). Local turn penalties can be added via RESTRICTION graph grammar identifiers over dummy edges (bootm).}
    \label{turns}
\end{figure}

Geospatial or property (social) graphs can easily be generated using our intuitive endpoints that can be used in R/C++/Java/JavaScript/Python API forms or in SQL syntax. Road network graphs are naturally geo based, however, it is completely possible to generate graphs over nano dimensional scales as well in Kinetica-Graph. Turn penalties can be added on demand, i.e., angle based turns can be plugged into an existing graph topology via a novel concept called dummy edges/nodes that is completely hidden to the user. However, adding these dummy entities enable us to solve network path problems without having to embed the combinatorial `if conditions' inside the solver algorithms. Hence this process is completely segregated from the solver design, which makes our solvers lean and efficient. It is also possible to add local penalties and restrictions with specific identifiers that can be set either at graph creation or solve time as depicted in Figure~\ref{turns}.

Multiple node and edge labels can be attached efficiently without any limits over the property graphs and utilized in solve and query endpoints. Graphs can in general be single, replicated or partitioned in Kinetica. Replicated and partitioned graphs use multiple graph servers in distributed cluster architectures using the efficient Pull/Push ZeroMQ~\cite{zmq} inter-processor communication pattern as discussed in Section~\ref{Section:communication}. Extensive set of parallel at-scale graph solvers are implemented in Kinetica that mostly use robust amd proven OpenMP technology~\cite{openmp}.

\begin{figure}
\centering
    \includegraphics[width=0.7\linewidth, keepaspectratio]{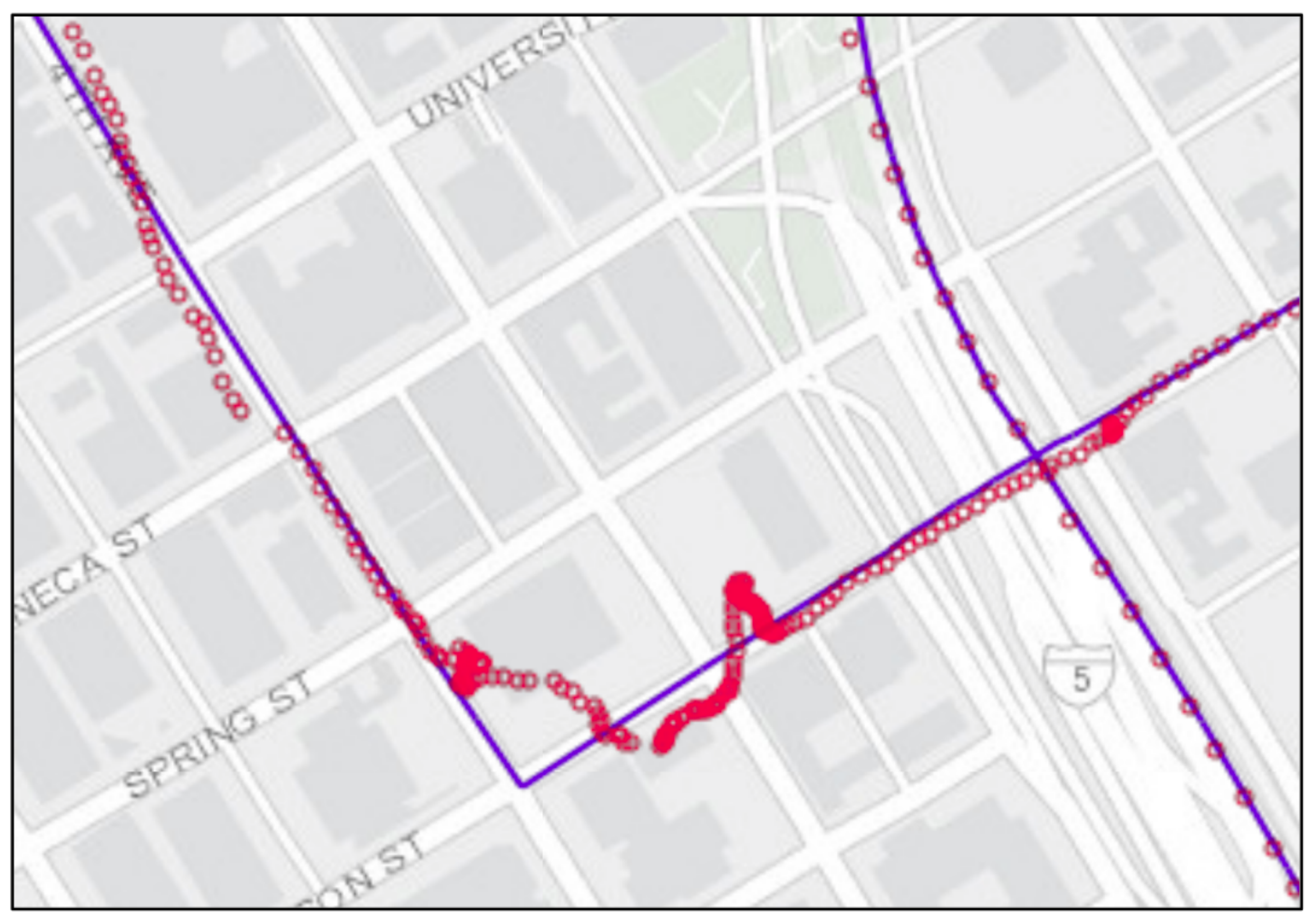}
    \caption{Noise in the GPS samples is quite visible by the
void red circles zoomed in on a portion of the Microsoft’s
Seattle data set. The Map matching algorithm finds the
best route by screening possible path sequences under the
constraints of the graph road network topology.~\cite{mapmatching}}
    \label{mapmatch}
\end{figure}

Map matching solver using hidden Markov chains deserves a special mention among many note-worthy Kinetica-Graph solvers since its success stems mainly from graph database's efficient doubly link topology structure explained in Section~\ref{Section:topology}. This patented in-house capability determines the route of thousands of GPS emitting vehicles using a novel adaptive width Hidden Markov Chain algorithm~\cite{mapmatching} shown in Figure~\ref{mapmatch}. On one test batch consisted of more than $300K$ sample points belonging to $370$ individual trips of varying degrees of sampling frequencies between 0.5 seconds and 5 seconds, we were able to obtain results in less than $24$ seconds using 8 cores where 95 percent of the trips had match scores well below 1 meter over a graph of approximately 7 million edges.

\begin{figure}
\centering
    \includegraphics[width=\linewidth, keepaspectratio]{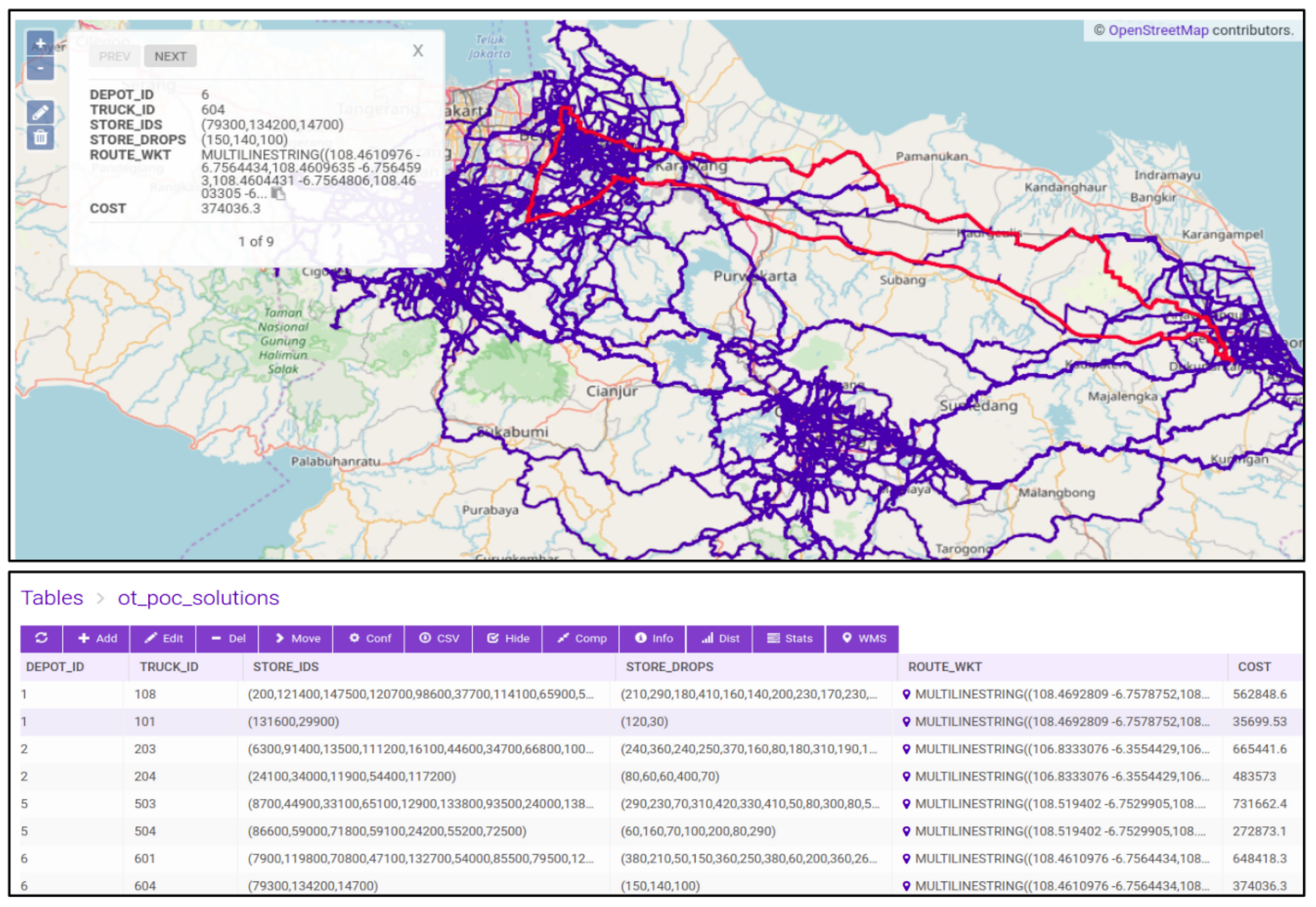}
    \caption{Truck $604$ started from depot $6$, stops at three customer locations delivering the respective amounts of $150$, $140$, and $100$ units and coming back to the same originating depot $6$ among $300$ trucks from $7$ depots in total. The output table shows the routing for all trucks with the respective delivery amounts (bottom).}
    \label{msdo}
\end{figure}

Another Match-Graph endpoint solver, a.k.a., MSDO (multiple supply demand optimization) enumerates millions of combinations in milliseconds and provides the dynamic routing and tracking capability for the entire distribution fleet to thousands of customer locations in the most optimal manner~\cite{msdoblog}. Three hundred ($300$) trucks emanating from seven depots all with varying capacities distribute over a set of three thousand and five hundred customer locations with varying sizes of demands. Our MSDO solver computes three hundred truck routes in the most optimal manner in less than two minutes (114 secs) using a multi-core (80 cores) single node platform over a geography of 400 miles across the Greater Jakarta region as shown in Figure~\ref{msdo}. The use of four graph servers in replicated graph mode, further lessens this already fast solve timing to a mere 45 seconds.

\begin{figure}
\centering
    \includegraphics[width=\linewidth, keepaspectratio]{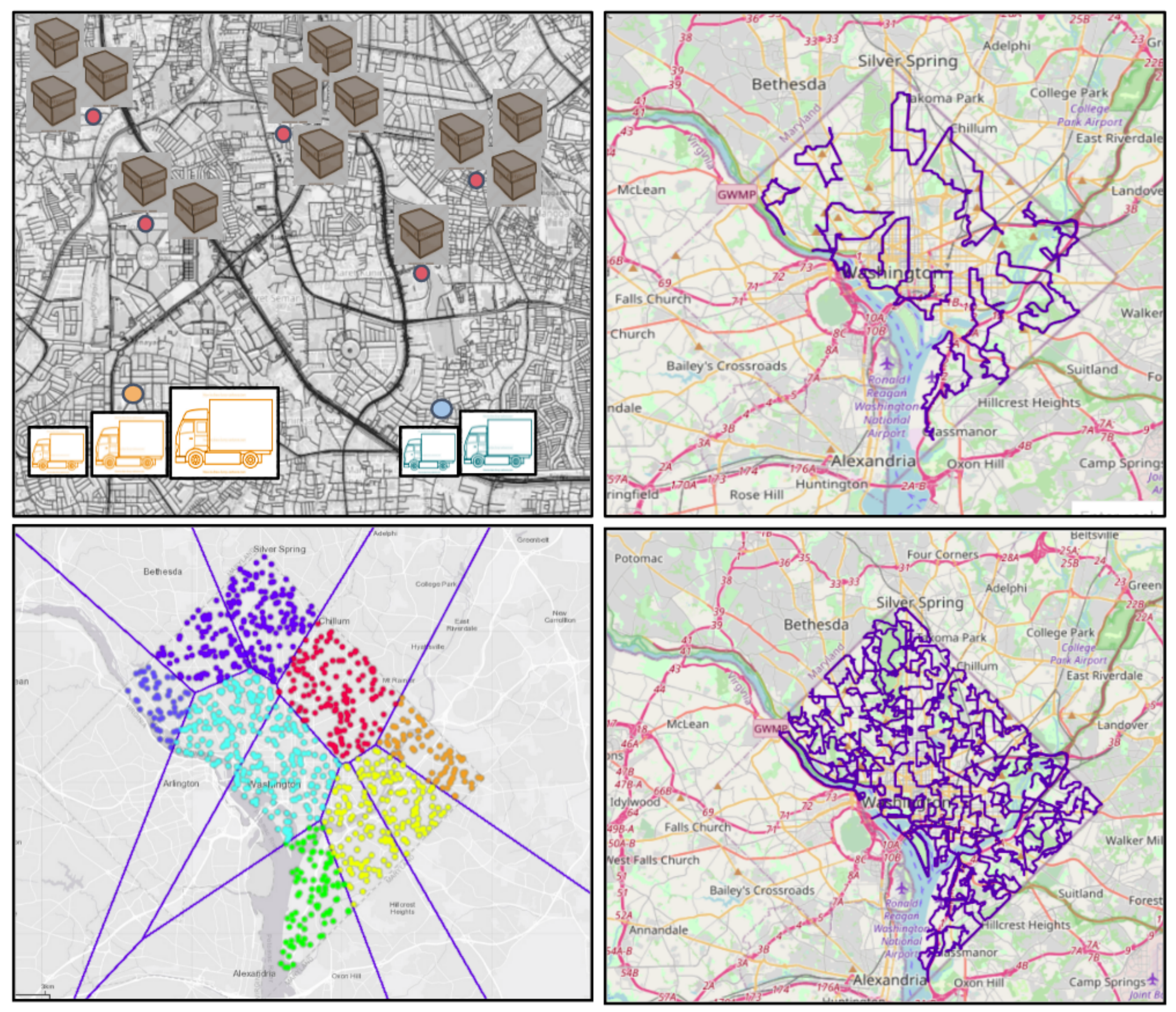}
    \caption{Application of MSDO to the problem of multiple traveling salesman (top-left); one thousand ($1000$) random locations in DC metropolitan area is distributed using Voronoi partitioning into ten ($10$) zones; each zone's generator point is considered to be a collector (supply) picking from collections in random locations (demands) shown in different colors (bottom-left). The problem is cast into MSDO format by considering one truck per collector of size equal to the number of collections within the closure of the collector's own Voronoi zone (top and bottom-right, respectively).}
    \label{msdodc}
\end{figure}

Another problem case is a classic multiple traveling salesman problem: One thousand random locations are generated and depicted as collection locations within the metropolitan region of Washington-DC. Kinetica’s ST\_voronoi geometry function is invoked followed by a geo-join operation to split and assign 1000 collections to 10 collectors (generator points in Voronoi partitioning are user-prescribed). The problem is to find the optimal round trips for each of these ten collectors. This classic multiple traveling salesman problem can be cast into MSDO format as if there is one truck at each collector of size equal to the number of collections they each need to visit. The results are shown in Figure~\ref{msdodc}.

\begin{figure}
\centering
    \includegraphics[width=\linewidth, keepaspectratio]{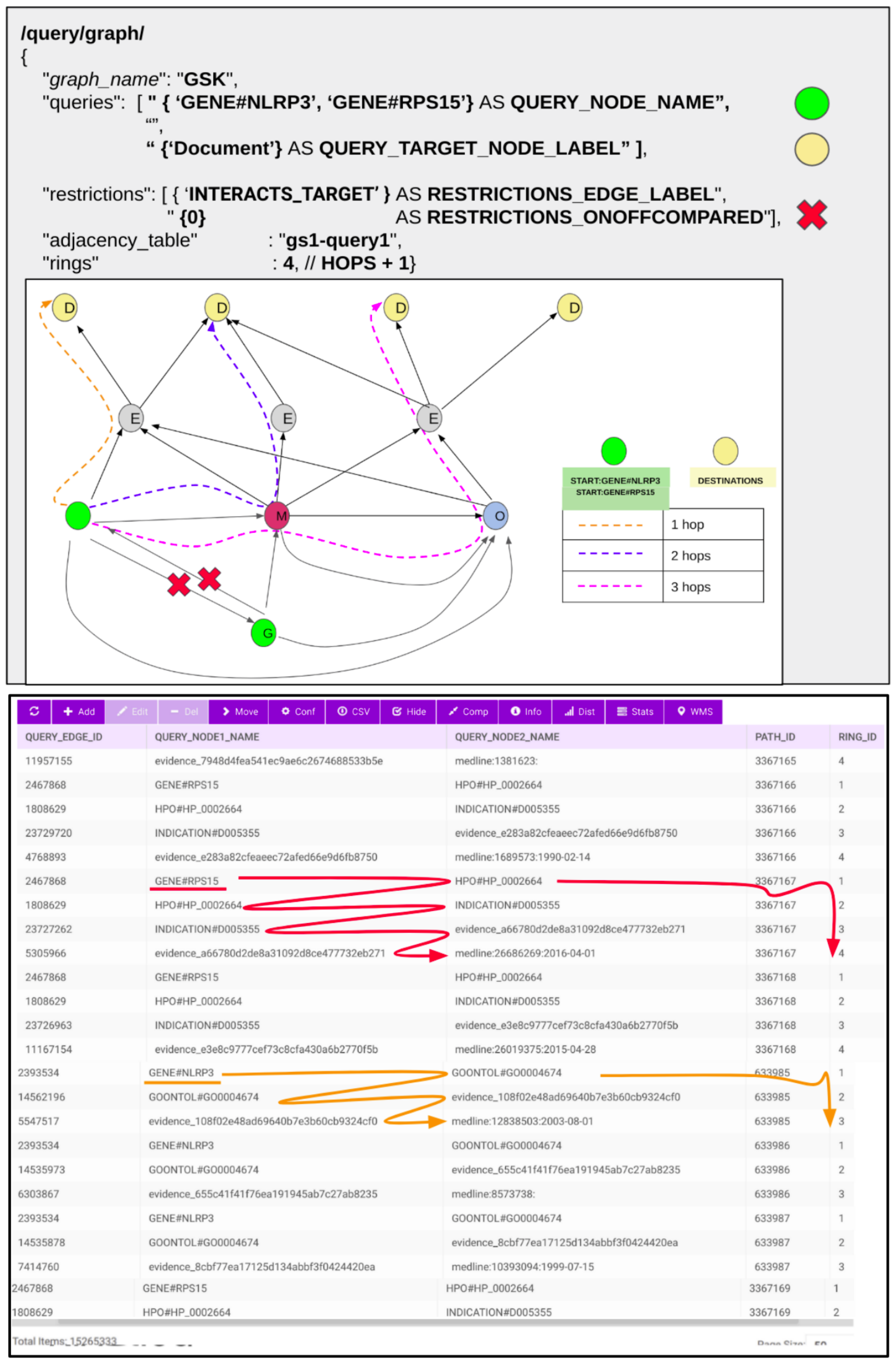}
    \caption{Finding all the document (D) paths via evidences (E) linked to either one of the genes NLRP3 and RPS15 by skipping the mutation links among the genes through observations (O) within three (3) hops (top). Query-Graph is able to find $\approx~240K$ paths in a few seconds using multiple graph servers (bottom).}
    \label{query}
\end{figure}

Kinetica-Graph’s powerful adjacency Query engine is capable of traversing millions of graph nodes starting from a set of nodes to a set of target nodes, i.e., many-to-many fashion with at-scale performance functionally compliant with the cypher language but instead using its own extendible and flexible graph grammar. As an example use-case from the pharmaceutical industry, our Graph-Query engine is able to find all the paths from a particular gene set via the relevant links to evidence based nodes leading  to `documentation' labeled nodes within three hops in a 27 million graph in matters of a few seconds for the large gene-evidence-document database as shown in Figure~\ref{query}.

\begin{figure}
\centering
    \includegraphics[width=\linewidth, keepaspectratio]{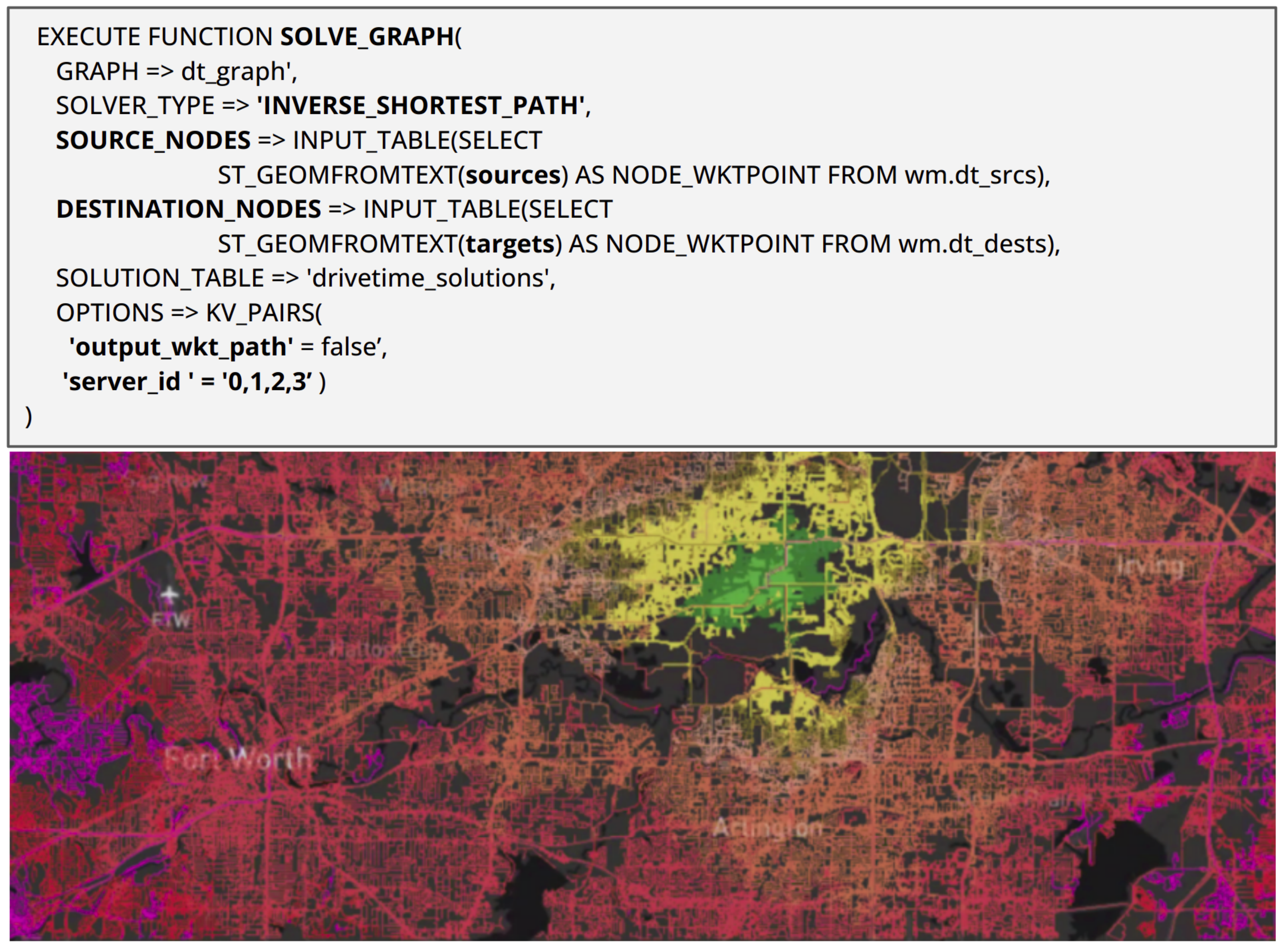}
    \caption{A distributed use case for solving the inverse shortest paths against one million random point pairs around Dallas-Fort Worth area. SQL Solve-Graph call (top), and the resulting WKT linestring paths shown in different colors with respect to the cost value by a WMS image visualization call (bottom). Four ($4$) Graph-Servers complete the task in $25$ seconds on a $8$ core machine of $24$ GBytes of memory.} 
    \label{batchsolve}
\end{figure}

Our batch solver, runs with at-scale and at-pace performance against $1$ million wkt point pairs on a 8 core laptop using 4 Graph-Servers under 25 seconds (6 minutes with all WKT linestring paths) as shown in Figure~\ref{batchsolve} with the accompanying Solve-Graph endpoint in SQL (each WKT linestrings paths have an average of $\approx~500$ chars).

\begin{figure*}
\centering
    \includegraphics[height=0.4\textheight, width=0.7\linewidth, keepaspectratio]{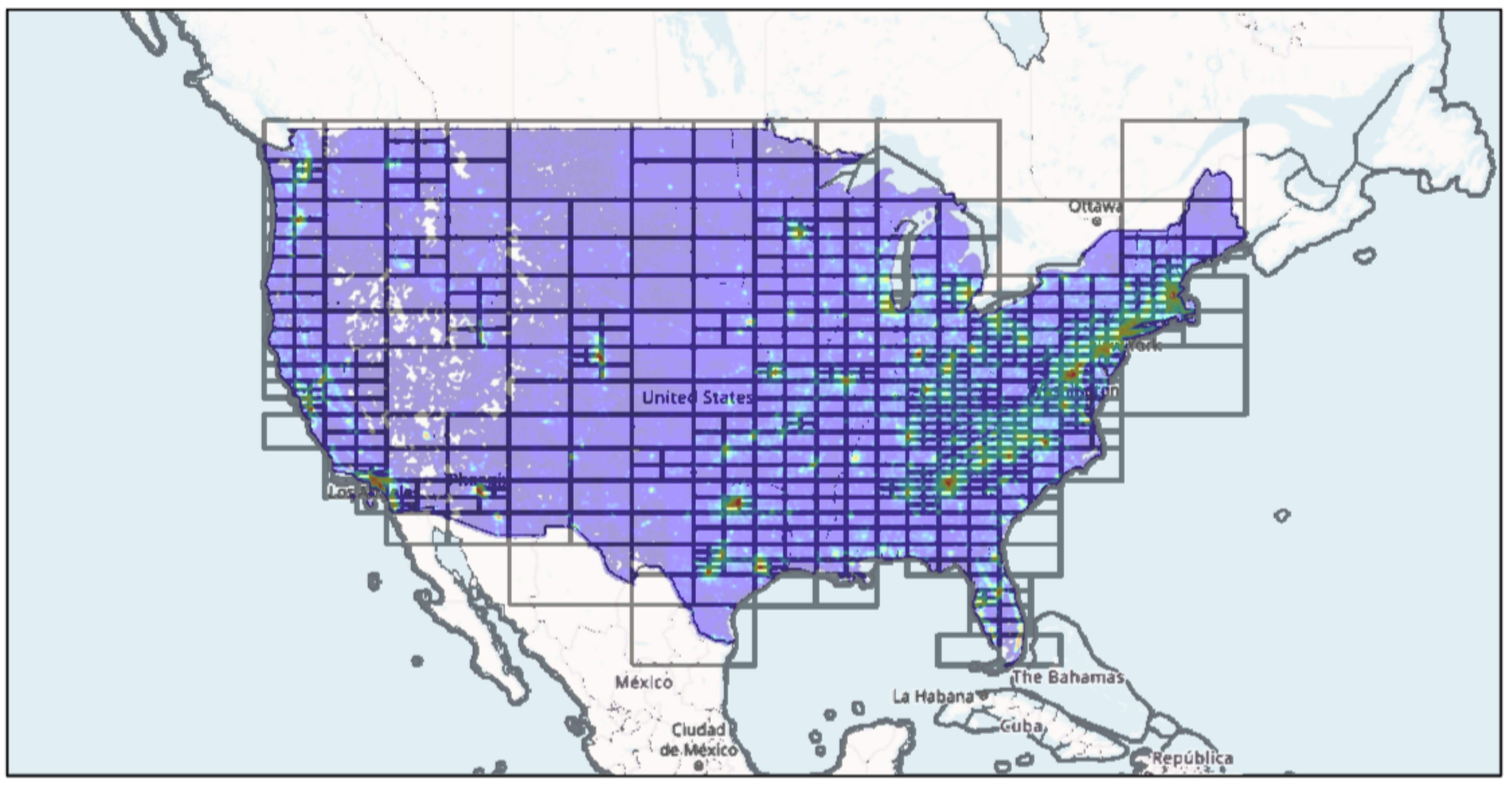}
    \caption{Adaptive division of US road network. OSM data is read and parsed into tiles that are adaptively divided when a tile reaches more than half a million edge records. There are approximately 1200 tiles in separate CSV files generated adaptively that corresponds to $\approx$260 million edges in total.}
    \label{world}
\end{figure*}

Finally, we have instrumented an automatic ingestion framework for our users to construct geo-based graphs given any arbitrary lon/lat WKT bounding box as input over the freely available Open Street Map database~\cite{osm} for the USA and its territories, otherwise a monumental task of generating over $260$ million edge roads including the service roads with weights set based on the legal speed limits. We have devised an adaptive splitting strategy like quad-tree but more flexible in that it can refine more than two-levels in adjacent quads, into separate CSV files ($1200$ in total) when the number of nodes within each quadrant becomes more than half a million as shown in Figure~\ref{world}. These files are stored in our file servers available for ingesting externally to Kinetica-DB and get updated periodically. When the user inputs a bounding box in lon/lat, we then find the intersecting quadrants and their corresponding CSV files to load from the storage into Kinetica-DB so that a single Create-Graph call could stitch the contents of multiple quadrants (CSV) properly to form one connected graph. Note that, the divisions are specifically constructed to result in non-overlapping edges shared by the quad tiles.

\begin{figure*}
\centering
    \includegraphics[height=0.4\textheight, width=0.7\linewidth, keepaspectratio]{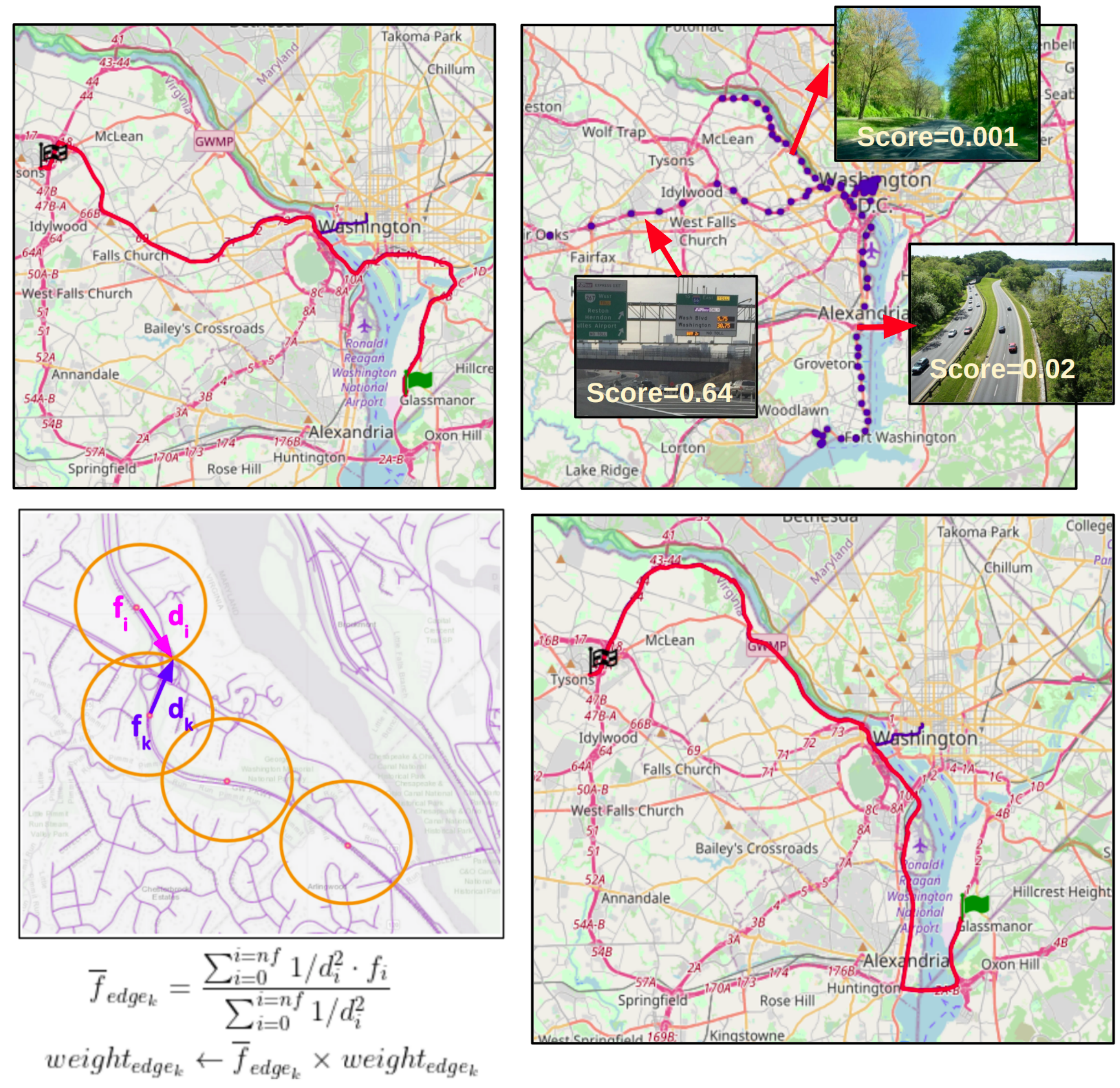}
    \caption{Scenic route versus shortest path using the scenic scores of a ML model. The shortest path (top-left). Solve-Graph call is passed a set of solve time WEIGHTS via the combination of \textit{WEIGHTS\_WKTPOINT, WEIGHTS\_FACTORSPECIFIED} where the locations are the hundreds of images around Washington DC with scenicity scores inferred from an ML model (scores close to zero if scenic otherwise one) (top-left). Internally the weights are interpolated by spreading the weight factor in an inverse distance weighted manner as shown in the formula (bottom-left). Solve-Graph using Dijkstra results in a scenic path due to the new weights favoring scenic routes (bottom-right).}
    \label{scenecity}
\end{figure*}

It is also worth noting that in general the scalar field of edge weights (impedances) can be modified spatially with appropriate identifiers such that the solve time weights could be imposed based on the changing traffic patterns. Similarly, any image input from an ML model can have an impact on the shortest paths. For example, a scenic route is computed instead of the shortest since the edge weights are overridden based on the scenic scores (scores close to zero when it is more scenic) computed by an ML model in Kinetica. Thousands of images with scene scores is spread over the graph network using an inverse distance weighted interpolative manner via the \textit{WEIGHTS} identifier, as shown in Figure~\ref{scenecity}.

There are countless applications of Kinetica-Graph, along with its hybrid distributed Kinetica-DB; we only covered a few use cases in this paper, however, the authors highly encourage the readers to download and try exercising the Kinetica-Graph endpoints under the guidance of hundreds of on-line tutorials and publicly accessible videos. Kinetica's Developer Edition is freely available here https://www.kinetica.com/try/.

Future works on Kinetica-Graph would most likely involve tighter integration with the ML models. Needless to say that we'd continue adding new at-scale parallel graph solvers into our Graph-Analytics stack to help increase the adoption of Kinetica-Graph. We'd also work on increasing the number of distributed algorithms for our many-to-many queries and non-Dijkstra solvers such as page rank and centrality between-ness using our robust many graph servers framework.

\section*{Acknowledgement}

The authors would like to thank the technical contributions of the entire Kinetica Engineering team, and more specifically, Vamshi Vangapalli for wrapping graph calls in SQL for Kinetica Workbench, Pat Khunachak for embedding graph into all Kinetica UIs,  Shouvik Bardhan for his invaluable know-how and advises and finally our CEO Nima Negahban for his strong support of Kinetica-Graph since its inception.

\section*{Notes on Contributors}
\small{
\noindent \textbf{Bilge Kaan Karamete} is the lead technologist for the Geospatial, Graph and Visualization efforts at Kinetica. His research interests include computational algorithm development, unstructured mesh generation, parallel graph solvers and computational geometry. He holds PhD in Engineering Sciences from the Middle East Technical University, Ankara Turkey, and post doctorate in Computational Sciences from Rensselaer Polytechnic Institute, Troy New York.

\noindent \textbf{Louai Adhami} is a principal engineer at Kinetica, and holds a PhD in robotics from INRIA. He works on high concurrency graph solvers and graphics capabilities. He enjoys doing software architecture for distributed systems and teaching at George Washington University, Washington DC.

\noindent \textbf{Eli Glaser} is VP of Engineering at Kinetica. He leads the development teams concentrating in data analytics, query capability and performance. Eli holds Master's in Electrical Engineering from The Johns Hopkins University, Baltimore Maryland.
}

\section{Software avaliability}

Kinetica's Developer Edition is freely available here https://www.kinetica.com/try/.

\section*{References}

\bibliography{graph}

\end{document}